\documentclass[a4paper,10pt]{article}
\usepackage{amsmath}
\usepackage{amsfonts}
\usepackage{graphicx}
\usepackage{dsfont}
\usepackage[english]{babel}
\usepackage[utf8]{inputenc}
\usepackage{verbatim}
\usepackage{hyperref}

\def\qed{\hfill \vrule height 7pt width 7pt depth 0pt
              \medskip}
\DeclareMathOperator{\argmax}{argmax}
\DeclareMathOperator{\argmin}{argmin}
\newcommand{\be}{\begin{equation}}

\newcommand{\ee}{\end{equation}}

\newcommand{\bea}{\begin{equation*}}

\newcommand{\eea}{\end{equation*}}

\newcommand{\su}{\mathbf{u}} 

\newcommand{\sy}{\mathbf{y}} 
\newcommand{\RU} {\mathbf{U}} 
\newcommand{\RX}{\mathbf X} 
\newcommand{\RY}{\mathbf Y} 
\newcommand{\RUE}{\widehat{\mathbf U}} 
\newcommand{\ue}{\widehat{u}} 
\newcommand{\xe}{\widehat{x}} 
\newcommand{\UE}{\widehat{U}} 
\newcommand{\XE}{\widehat{X}} 
\newcommand{\dec}{\mathcal D}
\newtheorem{assumption}{Assumption}
\newtheorem{theorem}{Theorem}
\newtheorem{definition}[theorem]{Definition}
\newtheorem{proposition}[theorem]{Proposition}
\newtheorem{lemma}[theorem]{Lemma}
\newtheorem{corollary}[theorem]{Corollary}

\newtheorem{remark}{Remark}

\def\Z{\mathds{Z}}
\def\N{\mathds{N}}
\def\E{\mathbb{E}}
\def\R{\mathds{R}}

\def \P{\mathbf{P}}
\def \Pmed{\mathbf{\overline{P}}}
\def \FF{\mathcal F} 

\def \FB{\mathcal B}
\def \X{\mathbf{X}}
\def \Prob{\mathrm{P}} 
\def \ep{\varepsilon}
\def \qmd{\overline{\mathrm q}_d}
\def \qmad{\overline{q}}
\def \qd{\mathrm q_d}
\def \qx{\mathrm q_x}
\def \CC{\mathrm C}
\def \erfc{\text{erfc}}
\begin{document}
\title{Deconvolution of linear systems with quantized input: an information theoretic viewpoint.}
\author{Fabio Fagnani         \and
        Sophie M. Fosson 
}



\maketitle

\begin{abstract}
In spite of the huge literature on deconvolution problems, very little is done for \emph{hybrid} contexts where signals are quantized.
In this paper we undertake an information theoretic approach to the deconvolution problem of a simple integrator with quantized binary
input and sampled noisy output. We recast it into a decoding problem and we propose and analyze (theoretically and numerically) some low
complexity on-line algorithms to achieve deconvolution.
\end{abstract}
\section{Introduction}
The deconvolution problem is ubiquitous in many scientific and
technological areas such as seismology, astrophysics, image
processing and medical applications
(see e.g. \cite{ban:97,ber:98,byr:82,jai:89,spa:96,sta:02}). Its most general formulation
is as follows. We consider a time horizon $T$ (possibly infinite),
a convolution kernel $\mathcal{K}(t)$ and the input/output system
\be\label{deconvolution_problem} x(t)=\int_0^t
\mathcal{K}(t-s)u(s)ds\ee (we implicitly assume that $\mathcal{K}$
and $u$ are s.t. the above integral makes sense). The problem is
to estimate the input $u$ from some noisy version $y$ of the
output $x$.

This is an instance of inverse problem: to see why the problem is
difficult we focus on the special case $\mathcal{K}=1$ which will
be the case considered throughout this paper. In this context,
(\ref{deconvolution_problem}) can be written as
\be\label{integral} \dot x(t)=u(t)\,,\quad x(0)=0\,.\ee Since the
operation of differentiation is not robust with respect to noise
perturbation, the reconstruction of $u$ from $y$ cannot be simply
done by differentiation. The goal is then to estimate $u$, using
the available information on $x$ and any a priori information on
$u$. Several procedures can be exploited to accomplish this task
and the choice is in general motivated by a suitable trade-off
between precision of the solution and complexity of the algorithm.

Classical algorithms due to Tikhonov \cite{tik:63,tik:77} are
based on a penalization technique and work off-line: the
estimation $\ue$ at any time depends on the whole signal $y(t)$
with $t\in [0,T]$. This is a significant drawback in on-line or
interactive data flows application where the delay in estimation
is required to remain bounded. Causal algorithms have been studied
in \cite{fag:02,fag:03}, where bounds on the error have been
obtained for the case of bounded noises and regularity assumptions
on the input signals $u$.

An outstanding problem is how to use possible side information
available on the input signal $u(t)$ on the above algorithms:
indeed, while functional, and more generally convex, constraints
can be incorporated in the above algorithms, things are quite less
clear for more general constraints. In this paper we focus on the
case when $u$ is known to be a piecewise constant signal with
values restricted to a fixed known finite discrete alphabet. This
turns out to be a significant issue in the context of hybrid
systems where continuous-time systems are driven by discrete
digital signals. Such constraints are clearly of a non-convex type
and is not obvious how to include them in classical
deconvolutional algorithms.

In this work we will undertake an information-theoretic approach
to causal deconvolution problems with sampled quantized inputs
introducing algorithms which reconstruct $u$ through a decoding
procedure. A key feature of these algorithms 
is that they present very low complexity structure, while they
exhibit performance quite close to the information theoretical
limit. The main mathematical results consist in a rigorous
analysis of the asymptotic performance of the proposed algorithms
employing tools from the ergodic theory of Markov Processes.

In Section 2 we will give all the mathematical details regarding
the deconvolution problem with quantized input signals. In
particular, we will link it to classical decoding problems and we
will study the possibility to use classical decoding techniques
for our purpose. In Section 3 we will develop a
couple of low complexity deconvolution algorithms comparing their
performance. Section 4 is the core of our paper: it is devoted to
a deep analysis of the proposed algorithms. Using Markov Processes ergodic theorems we will be able to give
theoretical results on their behavior in the
asymptotic regime (time range going to $\infty$).

We conclude now the introduction with notation and terminology to be used throughout the paper.
\subsection{Notation}\label{Notation_section}
Given a subset $A$ of a set $\Omega$, $\mathds{1}_A:\Omega\to\{0,1\}$ is the indicator function, defined by $\mathds{1}_A(x)=1$ if
$x\in A$ and $\mathds{1}_A(x)=0$ otherwise. Erfc indicates the complementary error function, defined by  
$\erfc(x)=\frac{2}{\sqrt{\pi}}\int_x^{+\infty}e^{-s}ds$ for any $x \in \R$. $\mathcal{B}(\Omega)$ indicates the Borel $\sigma$-algebra
of $\Omega$.

Capital letters will be used to name random variables (r.v.'s for
short), while boldface capital letters will be  vectors whose components are r.v.'s.

 $\Prob$ will be the probability on discrete r.v.'s, while  $f_{(\cdot)}$ the probability density function of continuous or hybrid (that
is,
involving both continuous and discrete events) r.v.'s. Instead, $\P$ will denote the transition probability matrix
of a Markov Chain (Section \ref{MC_section}) and $P(\cdot,\cdot)$ the transition  probability kernel of a Markov Process
(Section \ref{MP_section}). Finally, $\E$ will be the mean operator.

\section{Statement of the problem}
\subsection{The deconvolution problem}\label{SectionDecProblem}
In the following we stick to the system
(\ref{deconvolution_problem}) under the assumptions we make throughout this paragraph.
\begin{assumption}
The available
output signal is a noisy, sampled version of $x(t)$:
$$y_k=x_k+n_k$$
where $x_k=x(\tau k)$, $\tau >0$ being the constant sampling time, and $n_k$'s are realizations of independent, identically distributed
Gaussian variables $N_k$'s of
$0$ mean and variance $\sigma^2$. 
\end{assumption}
We will denote by
${\bf y}=(y_1,\dots, y_K)\in\R^K$ the
vector of all available measures  ($K=T/\tau$ is assumed to be an integer) and by $\sy_a^b=(y_a,y_{a+1},\dots,y_b)$
the available measures from time $a$ to time $b$, with $a,b \in \{1,\dots,K\}$, $a<b$.

A deconvolution algorithm consists in a function $$\Gamma:\R^K\to
\R^{[0,T]}.$$ $\ue=\Gamma({\bf y})$ is the estimated input and
in general it will not coincide with the true input $u$. What in
general we request is a bound on the error $u-\ue$ and some
consistency property: when the variance of the noise and the
sampling time go to $0$, the error should converge (in some
suitable sense) to $0$.

We say that a deconvolution algorithm $\Gamma$ is causal (with
delay $k_0\tau$. $k_0\in \N$) if there exists a sequence of functions
$\Gamma_k:\R^{k+k_0}\to\R^{[(k-1)\tau, k\tau[}$, where
$k=1,2,\dots$, such that
$$\Gamma(\sy)|_{t \in [(k-1)\tau, k\tau[}=\Gamma_k(\sy_1^{k+k_0})\,.$$
Such an algorithm  estimates the unknown
signal in the current time interval $[(k-1)\tau, k\tau[$ exploiting the past and present information $y_1,\dots,y_k$ along with a
possible bounded future information $y_{k+1},\dots,y_{k+k_0}$.

We now come to the assumptions on the input signals.
\begin{assumption}
There is a finite alphabet $\mathcal{U}\subset \R$ and we
consider signals of type \be u(t)=\sum_{k=0}^{K-1} u_k
\mathds{1}_{[k\tau,(k+1)\tau[}(t)~~~~~u_k \in \mathcal{U}\,.\ee
\end{assumption}
$u(t)$, with $t \in [0,T[$, is then completely determined by the
sequence of samples $u_0,u_1,\dots,u_{K-1}$. For simplicity we
assume the sampling time $\tau$ to be the same as in the output
and to have an exact synchronization in the sampling instants. The
output signals are now identified by samples $x_1,x_2,\dots,x_K \in
\mathcal{X}$, where $\mathcal{X}\subset \R$ is a suitable alphabet
(recall that we have fixed $x_0=0$). Of course, in principle, one
could still use the deconvolution algorithms in
\cite{fag:02,fag:03} or \cite{tik:63,tik:77}, however, there would
be no way to use inside the algorithm the a priori information on
the quantization of $u$. Instead we now show that, in this case,
our deconvolution problem can completely be recasted into a
discrete decoding problem. Notice indeed that the input/output
system is simply described by
\be\label{coding}
\left\{\begin{array}{rl}
&x_0=0 \\
&x_{k+1}=x_k+\tau u_k \text{,    } ~~~k=0,\dots,K-1.
\end{array}
\right.
\ee The
vector ${\bf x}=(x_1,\dots x_K)$ can thus be seen as a coded version
of ${\bf u}=(u_0,\dots, u_{K-1})$: we can write ${\bf x}={\cal
E}({\bf u})$ where $\cal E$ denotes the encoder given by (\ref{coding}). Afterwards, ${\bf x}$ is transformed as it was
transmitted through a classical Additive White Gaussian Noise
(AWGN) channel, the received output being given by $y_k=x_k+n_k$.

It is on the basis of these measures that we have to estimate the
'information signal' ${\bf u}$. Notice that the real time $t$ is
completely out of the problem at this point and everything can be
considered at the discrete sampling clock time. In the coding theory
language, a decoder is exactly a function $\dec:\R^K\to
{\mathcal U}^K$ which allows to construct an estimation of the input
signal: $\widehat\su=\dec({\bf y})$. Even in this context we can
talk about causal algorithm if there exists a sequence of functions
$\dec_k:\R^{k+k_0}\to {\mathcal U}$ such that
$$\dec(\sy)_{k-1}={\cal D}_k(\sy_1^{k+k_0})~~~~k=1,\dots K\,.$$
Finally,
\begin{assumption}
The unknown input is assumed to be generated by a stochastic source with a
known distribution, independent from the noise source.
\end{assumption}
The particular source distribution considered in this work will be introduced in Section \ref{inputassumptions}.

According to the notation given in Section \ref{Notation_section}, in the sequel $U_k$ will identify the input r.v. at time
$k$, $X_k$ the corresponding system output given by expression
\ref{coding}, $Y_k=X_k+N_k$ the measured output, $N_k$ being the
Gaussian noise. Furthermore, $\UE_k=\dec(\mathbf{Y})_{k}$ and $\XE_k=\XE_{k-1}+\tau \UE_{k-1}$ ($\XE_0=0$) will be respectively the
estimated
input and the estimated state. Finally,
$\RU=(U_0,\dots,U_{K-1})$, $\RUE=(\UE_0,\dots,\UE_{K-1})$, $\RY=(Y_1,\dots,Y_{K})$, $\RY_a^b=(Y_a,\dots,Y_b)$, $a,b \in \{1,\dots,K\}$,
$a<b$.

\subsection{Error Evaluation: The Mean Square Cost}
A fundamental issue in the deconvolution problem is the choice of
the norm with respect to which errors are evaluated. In this context, we consider the mean square cost:
\bea
\begin{split}
    \overline d(\dec)&=\tau\E \left(||\RU-\RUE||^2\right)=
 \tau \sum_{k=0}^{K-1}\E
\left(|U_k-\widehat{U}_k|^2\right).
\end{split}
\eea
We now define $\dec^*$ as the decoder minimizing $\overline d(\dec)$ among all
the possible decoders. It can be constructed as follows: given the density
$f_{\RY}(\sy)$ of $\RY$, notice that
\bea\overline d(\dec)=\tau\sum_{k=0}^{K-1}\int_{\R^K}\E\big(|U_k-\dec(\sy)_k|^2|\RY= \sy\big) f_{\RY}(\sy)d\sy.\eea
Hence, for any $\sy \in \R^K$,
\bea
\dec^*(\sy)_k=\underset{v \in \mathcal U}{\argmin}\,
\E\big(|U_k-v|^2|\RY= \sy\big)=\underset{v \in \mathcal U}{\argmin}\sum_{u \in \mathcal U}|u-v|^2\Prob(U_k=u|\RY=\sy).\eea
This turns out to be a finite optimization problem which can be solved by means of a
marginalization procedure and a Bayesian inversion: \bea
\Prob(U_k=u|\RY= \sy)=\sum_{\su \in \mathcal U^K: u_k=u}\frac{f_{(\RY|\RX)}
(\sy|\mathcal E(\su))\Prob(\RU=\su)}{f_{\RY}(\sy)}.
\eea
Analogously, we can define $\dec^{*_{k_0}}$ as the decoder minimizing $\overline d(\dec)$
among all the possible causal decoders with delay $k_0$:
\be\label{optimal_caus_dec} \dec^{*_{k_0}}(\sy)_{k-1}=\dec^{*_{k_0}}_k(\sy_1^{k+k_0})=
\underset{v \in \mathcal U}{\argmin}\sum_{u \in \mathcal U}|u-v|^2\Prob(U_{k-1}=u|\RY_1^{k+k_0}=\sy_1^{k+k_0}).\ee

\subsection{The BCJR algorithm}
In practice, the decoder $\dec^*$ can be implemented with the
well-known BCJR algorithm
\cite{bcjr:74}. This algorithm computes the probabilities of states
and transitions of a Markov source, given the observed channel
outputs; in other words, it provides the so-called APP (a posteriori
probabilities) on states and transitions,
therefore on coded and information symbols.

Let us briefly remind the BCJR procedure.
For $i,j \in \mathcal X$, we define the following probability density
functions: \be\label{def_alpha_beta}\begin{split} &\alpha_k(i)=f_{(X_k,\RY_1^k)}(i,
\sy_1^k)~~~~~~~~~~~~~~~~~~k=1,\dots,K\\
&\beta_k(i)=f_{(\RY_{k+1}^K|X_k)}(\sy_{k+1}^K|i)~~~~~~~~~~~~~k=0,\dots,K-1\\
&\Gamma_k(i,j)=f_{(X_k,Y_k|X_{k-1})}(j,y_k|i)~~~~~~~~k=1,\dots,K.
\end{split}\ee
For any $k=1,\dots,K$, the APP on states and on transitions respectively are:
\bea\begin{split} \lambda_k(i)&=f_{(X_k,\RY)}(i,\sy)~~~~~~~~~~~~
\sigma_k(i,j)=f_{(X_{k},X_{k-1},\RY)}(j, i, \sy).
\end{split}\eea
Given the following initial and final conditions:
\bea\begin{split}
&\alpha_0(i)=\Prob(X_0=i)=\left\{\begin{array}{rl}
1& \text{  if  }i=0 \\
0& \text{ otherwise.    }
\end{array}
\right.\\
&\beta_K(i)=1 \text{ for any } i \in \mathcal X
\end{split}\eea
 for $k=1,\dots,K$ we have
\bea\label{lambda}
\lambda_k(i)=\alpha_k(i)\beta_k(i)\eea \be\label{sigma}
\sigma_k(i,j)=\alpha_{k-1}(i)\Gamma_k(i,j)\beta_k(j) \ee where $\alpha_k(i)$ and $\beta_k(i)$,  $i\in \mathcal X$, can be respectively
computed with
a forward and a backward recursions: \be\label{forwbackwrec}
\alpha_k(i)=\sum_{h\in\mathcal{X}}\alpha_{k-1}(h)\Gamma_k(h,i)~~~~\beta_k(i)=\sum_{h\in\mathcal{X}} \Gamma_{k+1}(i,h)\beta_{k+1}(h).\ee
The APP are then recursively computed and finally used to decide on the transmitted input sequence.

Analogous causal versions of the BCJR algorithm can be used to
implement the decoder (\ref{optimal_caus_dec}) with delay $k_0$. For $k=1,\dots,K-k_0$, the APP on the transitions becomes
\be\label{CausalWithFutBits}\begin{split}
\widetilde{\sigma}_k(i,j)&=f_{(X_k,X_{k-1},\RY_1^{k+k_0})}(j,i,
\sy_1^{k+k_0})=\alpha_{k-1}(i)\Gamma_k(i,j)\widetilde{\beta}_k(j)\end{split}
\ee
where $\alpha_k$ and $\Gamma_k$ are defined as above, while $\widetilde{\beta}_k(j)=f_{(\RY_{k+1}^{k+k_0}|X_k)}(
\sy_{k+1}^{k+k_0}|j)$. For $k>K-k_0$, we recast into the
classical formulation (\ref{sigma}). For brevity, we will refer to the causal BCJR as CBCJR.
\subsection{Further Assumptions}\label{inputassumptions}
In the sequel of this work, we will make two further assumptions on the input:
\begin{assumption}\label{ass_bin}
The input alphabet is binary: $\mathcal U=\{0,1\}$. 
\end{assumption}
\begin{assumption}\label{ass_ber}
For $k=0\dots K-1$, the $U_k$'s are independent and uniformly distributed: $\Prob(U_k=0)=\Prob(U_k=1)=\frac{1}{2}$. In particular, the
$U_k$'s are independent from the Gaussian noises $N_k$'s.
\end{assumption}
Now the probabilistic setting introduced at the end of Section \ref{SectionDecProblem} is complete and we can  resume the system as
follows: given $X_0=\XE_0=0$, for $k=1,\dots, K$,
\be\label{probsett_resume}
\begin{split}
&U_{k-1}\sim \text{ Bernoulli }\left(1/2\right);\\ 
&X_k=X_{k-1}+\tau U_{k-1};\\
&N_k\sim\mathcal{N}(0,\sigma^2);\\
&Y_k= X_k+N_k;\\
&\UE_{k-1} =\dec(\mathbf{Y})_{k-1};\\
&\XE_{k}=\XE_{k-1}+\tau \UE_{k-1}.
\end{split}\ee
Notice that also $X_k$'s are independent from $N_k$'s.

Under Assumption \ref{ass_bin}, $$\overline d(\dec)=\tau
\sum_{k=0}^{K-1}\E \left(|U_k-\widehat{U}_k|\right)=\tau K
\Prob_b(e)$$ where \be\label{BER}
    \Prob_b(e)=\frac{1}{K}\sum_{k=0}^{K-1}\Prob(\widehat
    U_k \neq
    U_k)=\frac{1}{K}\E(|\RU-\RUE|)\ee
is the so-called Bit Error Rate (also denoted by BER), a very
common performance measure in digital transmissions that expresses the average number of bits in error. In our
context, minimizing $\overline d(\dec)$ is equivalent to minimizing
the BER and, therefore, the optimal decoder $\dec^*$ that performs
this minimization coincides with the well-known Bit-MAP (Maximum a posteriori)
decoder (see \cite{ric:08,bcjr:74}):\be\label{optimal_dec_2}
   \dec^*(\sy)_k= \underset{u \in \{0,1\}}{\argmax}~\Prob(U_k=u|\RY=
\mathbf y).\ee
Its causal version is given by
\be\label{optimal_caus_dec_2}
   \dec^{*_{k_0}}(\sy)_k= \underset{u \in \{0,1\}}{\argmax}~\Prob(U_k=u|\RY_1^{k+1+k_0}=
\mathbf y_1^{k+1+k_0}).\ee
We introduce here also the Conditional Bit Error Rate, CBER for short:
\be\label{BER_conditioned}
    \Prob_b(e|\RU)=\frac{1}{K}\sum_{k=0}^{K-1}\Prob(\widehat
    U_k \neq
    U_k|\RU)=\frac{1}{K}\E(|\RU-\RUE|\;{\mid} \RU).\ee
While the BER is a parameter that evaluates the
\emph{mean} performance of the transmission model, the CBER describes its behavior
for \emph{each} possible sent sequence. The CBER is then a relevant
parameter for our system, whose decoding performance changes in function of the transmitted input.

For computational simplicity, from now onwards let \be \tau=1 \ee so that $\mathcal X=\{0,\dots,K\}$ and in particular, if $X_0=0$, $X_k
\in \{0,\dots,k\}$.
In the BCJR implementation of decoders (\ref{optimal_dec_2}) and (\ref{optimal_caus_dec_2}), we obtain that
$\alpha_k(i)$, $i=0,1,\dots,K$, is null for
any $i>k$, while matrices $\Gamma_k$ and $\sigma_k$ are non-null
only on diagonal and superdiagonal.
By Assumption \ref{ass_ber},  $\Prob(X_k=j|X_{k-1}=i)=1/2$ if $j=i,i+1$ and
0 otherwise. Recalling that
the transition between $X_k$ and $Y_k$ is modeled by an AWGN
channel,
$f_{(Y_k|X_k)}(y_k|j)=\frac{1}{\sigma\sqrt{2\pi}}\exp\left(-\frac{(y_k-j)^2}{
2\sigma^2}\right)$,
we obtain \be\label{Lambda_AWGN}\begin{split}
\Gamma_k(i,j)&=f_{(Y_k|X_k)}(y_k|j)\Prob(X_k=j|X_{k-1}=i)\\
&=\frac{1}{2\sigma\sqrt{2\pi}}\exp\left(-\frac{(y_k-j)^2}{2\sigma^2}
\right)\;\;\;
\text{ for } j=i,i+1.\end{split}\ee
Given $\Gamma_k$, $\sigma_k$ or its causal version $\widetilde{\sigma}_k$ can be recursively computed
and the corresponding decoding rules are: \be\label{dec_rule}\text{BCJR}~~~~~~~~~~ \dec^*(\sy)_{k-1}= \left\{\begin{array}{rl}
0 &\text{  if  }
\sum_{i=0}^{k-1}\sigma_k(i,i+1)\leq\sum_{i=0}^{k-1}\sigma_k(i,i)\\
1&\text{  otherwise.  }
\end{array}
\right. \ee
\be\label{dec_rule_caus}\text{CBCJR}~~~~~~~~~~ \dec^{*_{k_0}}(\sy)_{k-1}= \left\{\begin{array}{rl}
0 &\text{  if  }
\sum_{i=0}^{k-1}\widetilde{\sigma}_k(i,i+1)\leq\sum_{i=0}^{k-1}\widetilde{\sigma}_k(i,i)\\
1&\text{  otherwise.  }
\end{array}
\right. \ee

\section{Suboptimal Causal Decoding Algorithms}
\begin{figure}
  \centering
  \includegraphics[width=7cm, viewport=60 50 770 550]{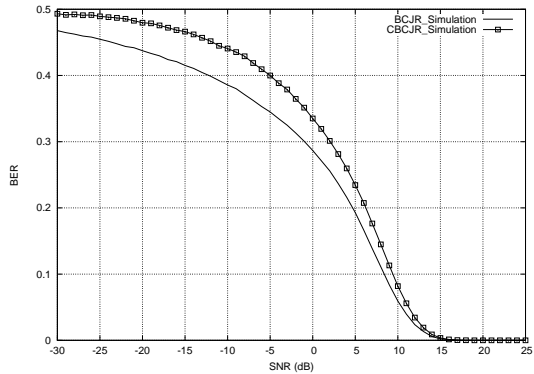}
  \caption{BCJR vs CBCJR.}\label{BCJRvsCBCJR}
\end{figure}
\begin{table}
\centering
\medskip
\begin{tabular}{l|l|l|l}
At step $k$: & Computations & Storage Locations & Decoding Delay \\
 \hline
  BCJR & $O(k)$ &$O(k)$  & $K-k$ \\
 \hline
 CBCJR & $O(k)$ &$O(k)$ & $k_0=0$ \\
\end{tabular}\caption{}\label{table_compare0}
\end{table}
Causality has a price  and the CBCJR algorithm has clearly a worse
performance than BCJR.

By simulating our system, we quantify the performance gap between
BCJR and CBCJR ($k_0=0$) as we can appreciate in Figure
\ref{BCJRvsCBCJR}: the two curves represent the corresponding
BER's in function of the Signal-to-Noise Ratio (SNR), here defined
as $\tau^2/\sigma^2=1/\sigma^2$. These outcomes are the averages over 5000 transmissions, each of which being a 100 bit message.
avoid unacceptable delays and complexity problems in the BCJR and CBCJR implementation).
We remark that CBCJR has the best performance among causal deconvolution algorithms.

Moreover, by comparing the efficiency of the two procedures (the
results are reported in Table \ref{table_compare0}), we gather
that for both BCJR and CBCJR the required computations and storage
locations linearly increase with the number of transmitted bits,
which is a drawback in case of long transmission.

 This fact motivates the development of new suboptimal causal algorithms that
improve the efficiency without substantial loss of reliability. To
achieve that, we implement the CBCJR fixing the number of states,
that is, at each step we save the $n$ states with largest
probability (where $n$ is arbitrarily chosen) and we discard the
others.

We now introduce the algorithms in the cases $n=1$ and $n=2$, which are of great interest for their low complexity, and we show some
simulations' outcomes.
\subsection{One State Algorithm}
A suboptimal causal decoder $\dec^{(1)}:\R^K\to\{0,1\}^K$ can be derived from the CBCJR by assuming the most probable state to be the
correct one. At any step $k=0,1,\dots$, $\dec^{(1)}$ decides on the current bit by a single MAP procedure and upgrades the estimated
state, which is the only one value that requires to be stored. 

Consider (\ref{def_alpha_beta}), (\ref{CausalWithFutBits}) and (\ref{dec_rule_caus}). Given the estimated state $\xe_{k-1}$, the
decoding rule of $\dec^{(1)}$ at time step $k$ is given by (\ref{dec_rule_caus}) with no backward recursion $\widetilde{\beta}_k(j)$ and
$\alpha_{k-1}(\xe_{k-1})=1$, $\alpha_{k-1}(j)=0$ for any $j\neq \xe_{k-1}$. This reduces the decoding task to the comparison between two
distances; in fact, the One State algorithm that implements $\dec^{(1)}$ is as follows: 
\begin{enumerate}
    \item Initialization: $\xe_0=0$;
    \item For $k=1,\dots,K$, given the received symbol $y_k \in
    \mathds{R}$,
     \be\label{algoOne}\begin{split}&\ue_{k-1}=\dec^{(1)}(\sy)_{k-1}={\underset{u \in
\{0,1\}}{\argmax}}~\Prob(U_{k-1}=u|Y_k=y_k,X_{k-1}=\xe_{k-1})\\&~~~=
\left\{\begin{array}{rl}
0 &\text{  if  } \Gamma_k(\xe_{k-1},\xe_{k-1})\geq \Gamma_k(\xe_{k-1},\xe_{k-1}+1)\\
1&\text{  otherwise  }
\end{array}
\right.\\
&\xe_{k}=\xe_{k-1}+\ue_{k-1}\end{split}\ee
\end{enumerate}
and given the equality (\ref{Lambda_AWGN}) in the AWGN case, 
\be \Gamma_k(\xe_{k-1},\xe_{k-1})\geq \Gamma_k(\xe_{k-1},\xe_{k-1}+1)~~\Leftrightarrow~~|y_k-\xe_{k-1}|\leq |y_k-(\xe_{k-1}+1)|.\ee
\subsection{Two States Algorithm}\label{par_2states}
\begin{figure}
  \centering
  \includegraphics[width=6.5cm, viewport=70 0 650 470]{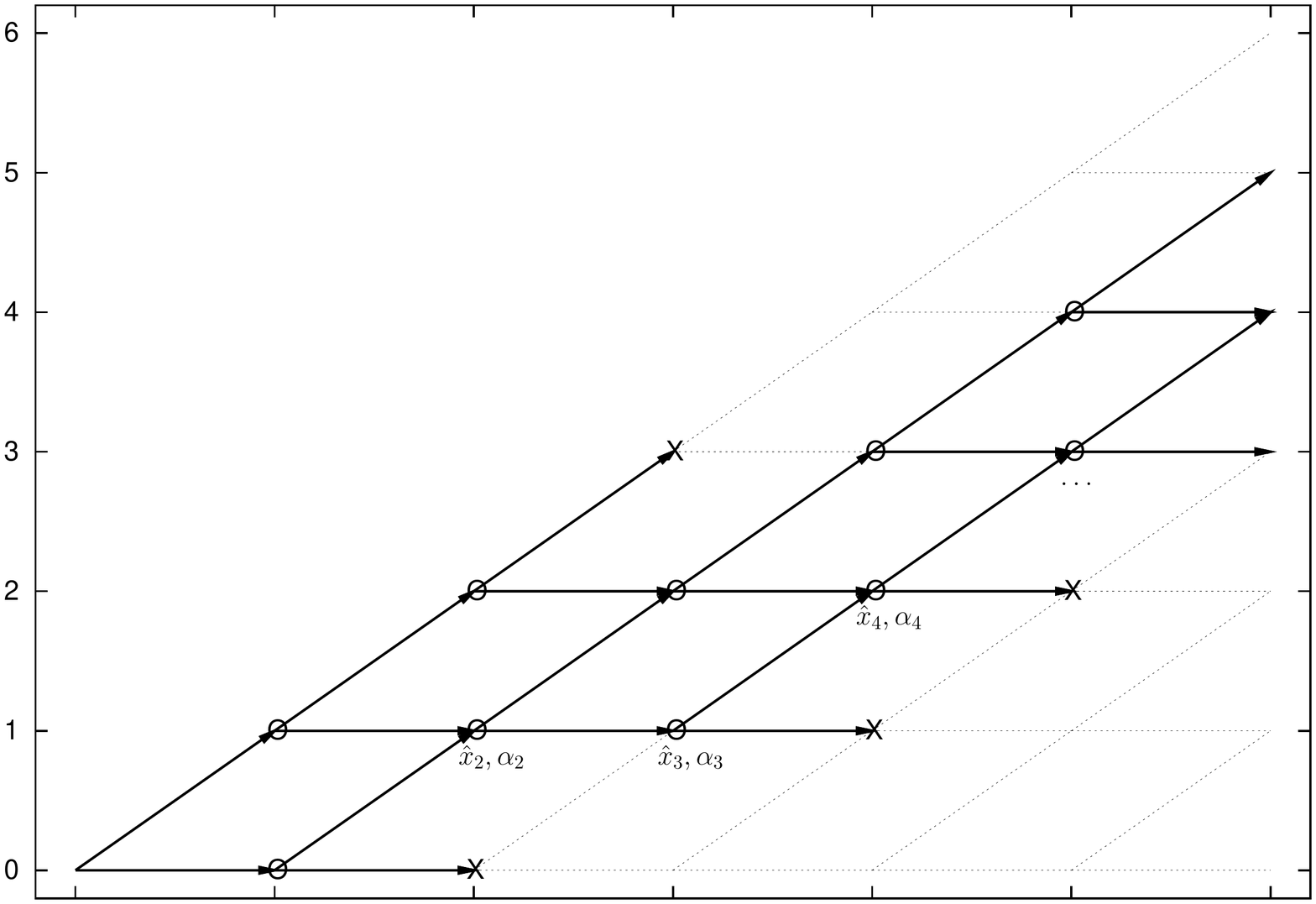}
  \caption{Trellis representation of the Two States Algorithm.}\label{BER_survive_trellis}
\end{figure}

By fixing $n=2$, we derive a decoder $\dec^{(2)}:\R^K\to\{0,1\}^K$ that, at each step, estimates the current input bit and computes and
stores the two most likely states along with the corresponding probabilities $\alpha_k(i)$ (defined by (\ref{def_alpha_beta})). As for
the One State Algorithm, the estimation of the input bit is performed by a MAP decoding rule (\ref{dec_rule_caus}) with no backward
recursion and summing over the two ``surviving'' states. In detail, the recursive Two States algorithm that implements $\dec^{(2)}$ is
the following:
\begin{enumerate}
 \item For $k=1$, given the unique starting state $\xe_{0}=0$, we estimate the first bit by a One State procedure:
 \be\begin{split}\ue_0=\dec^{(2)}(\sy)_{0}&={\underset{u \in \{0,1\}}{\argmax}}~\Prob(U_{0}=u|Y_1=y_1,X_0=0)\\&~~~=
\left\{\begin{array}{rl}
0 &\text{  if  } |y_1|\leq |y_k-1|\\
1&\text{  otherwise.  }
\end{array}
\right.
\end{split}\ee
Afterwards, the possible states are two: $\xe_1(0)=0$ and $\xe_1(1)=1$ and the corresponding probabilities
$\alpha_1(0)$ and $\alpha_1(1)$ in our framework are given by
\bea\begin{split}\alpha_1(j)&=f_{(X_1,Y_1)}(j,y_1)=f_{(Y_1|X_1)}(y_1|j)\Prob(X_1=j)\\&=f_{(Y_1|X_1)}(y_1|j)\Prob(U_0=j)=\frac{1}{2}f_{
(Y_1|X_1)}(y_1|j),~~~~ j\in\{0,1\}.\end{split}\eea
We then normalize these probabilities so that $\alpha_1(0)+\alpha_1(1)=1$ and we just store the couple of values
$(\alpha_1(0),\xe_1(0))$, as this is sufficient to retrieve also $(\alpha_1(1),\xe_1(1))=(1-\alpha_1(0),\xe_1(0)+1)$. For notational
simplicity we rename the stored vector  $(\alpha_1(0),\xe_1(0))$ as  $(\alpha_1,\xe_1)$. 
\item For $k=2,3,\dots,K$, given $(\alpha_{k-1},\xe_{k-1})$ and $F_k=f_{(X_{k},Y_1^{k})}(\xe_{k},y_1^{k})$
\bea\begin{split}
&\ue_{k-1}=\dec^{(2)}(\sy)_{k-1}=\\
&={\underset{u \in \{0,1\}}{\argmax}}~\Prob\big(U_{k-1}=u|Y_k=y_k, X_{k-1}=\xe_{k-1},F_{k-1}=\alpha_{k-1}\big)=\\
&=\left\{\begin{array}{rl}
0 &\text{  if  } 
\alpha_{k-1}\Gamma_k(\xe_{k-1},\xe_{k-1})+(1-\alpha_{k-1})\Gamma_k(\xe_{k-1}+1,\xe_{k-1}+1)\geq\\&\geq\alpha_{k-1}\Gamma_k(\xe_{k-1},
\xe_{k-1}+1)+(1-\alpha_{k-1})\Gamma_k(\xe_{k-1}+1,\xe_{k-1}+2)\\
1&\text{  otherwise.  }
\end{array}
\right.
\end{split}\eea
 From step $k-1$, three possible states arise: $\xe_{k-1}$, $\xe_{k-1}+1$ and $\xe_{k-1}+2$, whose probabilities are given by the
forward recursion in (\ref{forwbackwrec}):
    \be\label{zeta_i}\begin{split}
    &\alpha_k(\xe_{k-1})= \alpha_{k-1}\Gamma_k(\xe_{k-1},\xe_{k-1})\\
    &\alpha_k(\xe_{k-1}+1)= \alpha_{k-1}\Gamma_k(\xe_{k-1},\xe_{k-1}+1)+(1- \alpha_{k-1})\Gamma_k(\xe_{k-1}+1,\xe_{k-1}+1)\\
    &\alpha_k(\xe_{k-1}+2)=(1-\alpha_{k-1})\Gamma_k(\xe_{k-1}+1,\xe_{k-1}+2).
    \end{split}\ee
which can be reduced as follows in the case (\ref{Lambda_AWGN}):
  \bea\begin{split}
    &\alpha_k(\xe_{k-1})= \alpha_{k-1}\frac{1}{2\sigma\sqrt{2\pi}}\exp\left(-\frac{(y_k-\xe_{k-1})^2}{2\sigma^2}\right)\\
    &\alpha_k(\xe_{k-1}+1)=\frac{1}{2\sigma\sqrt{2\pi}}\exp\left(-\frac{(y_k-(\xe_{k-1}+1))^2}{2\sigma^2}\right)\\
    &\alpha_k(\xe_{k-1}+2)=(1-\alpha_{k-1})\frac{1}{2\sigma\sqrt{2\pi}}\exp\left(-\frac{(y_k-(\xe_{k-1}+2))^2}{2\sigma^2}\right).\\
    \end{split}\eea

Since  $|y_k-(\xe_{k-1}+1)|\neq \max\{|y_k-(\xe_{k-1}+j)|,j=0,1,2\}$, in the AWGN case $\alpha_k(\xe_{k-1}+1)\neq
\min\{\alpha_k(\xe_{k-1}+j),j=0,1,2\}$. Hence, the state $\xe_{k-1}+1$ is never discarded and also the two ``surviving'' states are
always adjacent. Therefore,
\begin{itemize}
 \item we calculate $\alpha_{\min}=\min\{\alpha_k(\xe_{k-1}),\alpha_k(\xe_{k-1}+2)\}$.
\item If $\alpha_{\min}=\alpha_k(\xe_{k-1})$, the surviving states are $(\xe_{k-1}+1,\xe_{k-1}+2)$ with probabilities
$(\alpha_k(\xe_{k-1}+1),\alpha_k(\xe_{k-1}+2))$. We then store the lowest state along with the corresponding normalized probability:
$(\alpha_{k},\xe_{k})=(\frac{\alpha_k(\xe_{k-1}+1)}{\alpha_k(\xe_{k-1}+1)+\alpha_k(\xe_{k-1}+2)},\xe_{k-1}+1)$.
\item Similarly, if $\alpha_{\min}=\alpha_k(\xe_{k-1}+2)$,
$(\alpha_{k},\xe_{k})=(\frac{\alpha_k(\xe_{k-1})}{\alpha_k(\xe_{k-1})+\alpha_k(\xe_{k-1}+1)},\xe_{k-1})$.
\end{itemize}
\end{enumerate}
\begin{remark}\label{extreme} When the extreme case $\alpha_k=1$ occurs, $\xe_{k}+1$ has null probability, then  $\xe_{k+1}=\xe_{k}$; 
analogously, when  $\alpha_k=0$, $\xe_{k+1}=\xe_{k}+1$. In these cases the Two States Algorithm actually behaves as the One State
Algorithm. \end{remark} 
\begin{remark}\label{initialstate}
As a consequence of Remark \ref{extreme}, the unique initial state $\xe_0=0$ can be interpreted as a double state with all the
probability in  $\xe_0=0$, that is, $(\alpha_0,\xe_0)=(1,0)$.
\end{remark}

\subsection{Simulations and comparisons}
\begin{figure}[h]
  \centering
  \includegraphics[width=9cm, viewport=60 50 770 550]{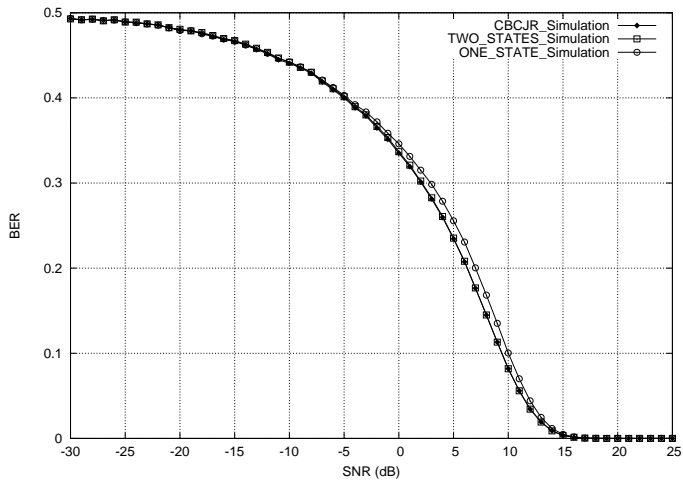}
  \caption{Performance comparison of different causal decoders.}
\label{BER_survive}
\end{figure} We report now the simulations' outcomes concerning the decoders $\dec^{*_0}$, $\dec^{(1)}$ and $\dec^{(2)}$, respectively
implemented with CBCJR, One State and Two States algorithms. The simulations have been performed considering 5000 different
transmissions, each of which being a 100 bit message.  The obtained results are then the averages overall transmissions. 

In Figure \ref{BER_survive} we compare the efficiency of the three decoding schemes, in terms of BER:  we evidence that two states are
sufficient to achieve performance very close to the causal optimum: we observe that the gain between $\dec^{(2)}$ and $\dec^{*_0}$ never
exceeds $0.15$ dB, while it achieves $0.8$ dB between $\dec^{(1)}$ and $\dec^{*_0}$ for BER's values between $0.2$ and $0.3$.
\begin{table}
\centering
\begin{tabular}{l|l|l|l}
At step $k$: & Computations & Storage Locations & Decoding Delay \\
 \hline
  BCJR & $O(k)$ &$O(k)$  & $K-k$ \\
 \hline
 CBCJR & $O(k)$ &$O(k)$ & $0$\\
\hline
 ONE STATE & $O(1)$ &$1$ & $0$ \\
\hline
 TWO STATES & $O(1)$ &$2$ & $0$ \\
\end{tabular}\caption{}\label{table_compare}
\end{table}
Moreover, as we report in Table \ref{table_compare}, the complexity of One State and Two States algorithms is constant when the number
is constant and no delay is produced in the decoding: this makes them efficient even for long-time transmissions, i.e., for a large
number of states.
\section{Suboptimal Causal Decoding Algorithms: Theoretic Analysis}
In this section, we propose an exhaustive theoretic analysis of One State and Two States algorithms and we provide a formal setting for
the analytical computation of their performance. According to
Definitions \ref{BER} and \ref{BER_conditioned} in Section 1, we will compute both the BER and the CBER, which respectively describe the
decoding for the ``mean input'' and for each possible input.

The natural setting of this analysis is the theory of Markov Processes, in countably infinite or not countable spaces (we will talk
about Markov Chains when the space is countably infinite).
\subsection{Theoretic Analysis of the One State Algorithm}
Suppose to transmit $K$ (possibly infinite) bits and
to decode by the One State method. The starting point of our analysis
is the definition, at any step $k=1,2,3\dots$, of the r.v.
\be\label{D_k} D_k=\XE_k-X_k\in\Z\ee $\XE_k$ being defined (\ref{probsett_resume}). $D_k$ actually represents the difference between the
actual and the estimated state values. Since $D_0=0$, the
following recursive relationship holds: \be\label{D_k_rec}
D_{k+1}=D_k+\UE_k-U_k\ee where $\UE_{k-1}=\dec^{(1)}(\sy)_{k-1}$ (see the algorithm (\ref{algoOne}).
While $U_k$'s are independent, $\UE_k$ is function of $U_k$ and $D_k$.
Then, the stochastic process $(D_k)_{k \in \N}$ is a Markov Chain (whose definition is formally given in the next section), which can be
exploited to carry on our analysis; in order to do that, let us first review some basic elements of Markov theory.
\subsubsection{Markov Chains}\label{MC_section}
The definitions and results introduced in this Section can be retrieved in the Chapter 3 of \cite{str:05} or in the Chapter 3 of
\cite{ler:03}.

By Markov Chain we intend any sequence of random variables $(X_n)_{n=0,1,\dots}$ assuming values in a countable set $\X$ and satisfying
the Markov property: $\Prob(X_{n+1}=y|X_{n}=x,X_{n-1},\dots,X_0)=\Prob(X_{n+1}=y|X_{n}=x)$. If the chain is time-homogeneous, that is
$\Prob(X_{n+1}=y|X_{n}=x)=\Prob(X_{n+m+1}=y|X_{n+m}=x)$, the \textit{transition probabilities}  $\Prob_{x,y}=\Prob(X_{n+1}=y|X_{n}=x)$
are the entries of the stochastic \textit{transition probability matrix} $\P \in [0,1]^{\X\times\X}$.

We review some important properties of a Markov Chain $(X_n)_{n=0,1,\dots}$ on $\X=\Z$:
\begin{definition}\cite[Section 3.1]{str:05}
Two states $x,y \in \Z$ \emph{communicate} if there exist $n,m \in \N$ s.t. $(\P^n)_{x,y}>0$ and
$(\P^m)_{y,x}>0$. If all the states communicate, the Markov Chain is said to be \emph{irreducible}.
\end{definition}
\begin{definition}\cite[Section 3.2.3]{str:05}
Let $\tau_j=\min\{n>0:X_n=j\}$: a state $j$ is said to be
\emph{positive recurrent} if $\E(\tau_j|X_0=j)<\infty$.  The Markov Chain itself is said to be positive recurrent if all its states are
so.
\end{definition}
\begin{proposition}\cite[Last part of Section 3.2.3]{str:05}
If a Markov Chain is irreducible and has one positive recurrent state, then all the states are so, that is the chain is positive
recurrent.
\end{proposition}

\begin{definition}\cite[Section 3.2.3]{str:05}
A invariant (or stationary) probability vector is a probability vector $\Phi$ (that is, $\Phi \in [0,1]^{\X}$ and $\sum_{x \in
X}\Phi_x=1$) such that $\Phi^T\P=\Phi^T$.
\end{definition}
The existence of an invariant probability vector, assured under some conditions, gives an important convergence result, as stated in the
following
\begin{proposition}\label{irr+posrec}\cite[Sections 3.2.3-3.2.4]{str:05} An irreducible, positive recurrent Markov Chain admits a unique
invariant probability vector $\Phi$. Moreover, $\Phi$ is the limit
of the so-called Ces\`{a}ro sum, that is
\bea \lim_{K\to\infty }\frac{1}{K}\sum_{k=0}^{K-1}(\P^k)_{x,d}=\Phi_d\;\;\;\forall x \in \Z.\eea
\end{proposition}
\subsubsection{The mean BER}
Let us go back to the One State algorithm. According to \ref{D_k_rec}, $(D_k)_{k \in \N}$ is a countable
homogeneous Markov Chain on $\Z$, with transition
probabilities \bea
\Pmed_{x,y}=\Prob(D_{k+1}=y|D_k=x)=\frac{1}{2}[\P_{x,y}(0)+\P_{x,y}(1)]\eea where $ \P_{x,y}(u)=\Prob(D_{k+1}=y|D_k=x,U_k=u)$, $u\in
\{0,1\}$.
Notice that the only non-null entries of $\P(u)$ are the following: \bea\begin{split}
&\P_{d,d+1}(0)=\frac{1}{2}\text{ erfc}
\left(\frac{d+\frac{1}{2}}{\sqrt{2}
\sigma}\right)~~~~\P_{d,d}(0)=1-\P_{d,d+1}(0)\\
&\P_{d,d}(1)=\frac{1}{2}\text{ erfc}
\left(\frac{d-\frac{1}{2}}{\sqrt{2}
\sigma}\right)~~~~~~~\P_{d,d-1}(1)=1-\P_{d,d}(1)
\end{split}\eea
$\Pmed$ is tridiagonal and, for any $x,y \in \Z$,
$\Pmed_{x,y}=\Pmed_{-x,-y}$ and $\Pmed_{x,y}>0$ if and only if $|x-y|\leq 1$; by iteration, for any $n \in \N$, $(\Pmed^n)_{x,y}>0$ if
and only if $|x-y|\leq n$. Hence, given any couple of states $x,y \in \Z$ with distance $|x-y|=m$, $(\Pmed^m)_{x,y}>0$ and
$(\Pmed^m)_{x,y}>0$, that is, $(D_k)_{k \in \N}$ is irreducible.
Moreover,
\begin{lemma}
$(D_k)_{k \in \N}$ is positive recurrent.\end{lemma}
\begin{proof} It suffices to apply the following criterion
proposed in \cite{str:05}: if there exists a function $\mathbf g \in
\R^{+\Z}$ so that $\mathbf g_x\geq (\Pmed
\mathbf{g})_{x}+\varepsilon$ for any $x \in \Z\setminus\{y\}$ and
for some $\varepsilon>0$, then $y$ is a positive
recurrent state.

In our case, it is easy to prove that $y=0$ is a positive recurrent
state considering $\mathbf g_x=|x|$. Moreover, given that the chain
is irreducible, if one state is positive recurrent, all states
are so.\qed\end{proof}
\begin{proposition}\label{inv_prob_vector} The following statements hold:
\begin{enumerate}
    \item $(D_k)_{k \in \N}$ admits a unique invariant probability vector $\Phi$;
    \item $\Phi$ is defined by
\be\label{inv_rel}\Phi_d=\Phi_0\prod_{i=1}^{|d|}
\frac{\Pmed_{i-1,i}}{\Pmed_{i,i-1}}
\ee where $\Phi_0=\left[1+2\sum_{d=1}^{\infty}\prod_{i=1}^{|d|}
\Pmed_{i-1,i}/\Pmed_{i,i-1}\right]^{-1}$.
\end{enumerate}
\end{proposition}
\begin{proof}
(1) It follows from Proposition \ref{irr+posrec}.\\
(2) By $(\Phi^T\Pmed)_d=\Phi^T_d$, for any $d \in \Z$, it follows
that \be\label{c_costant}
\Phi_{d-1}\Pmed_{d-1,d}-\Phi_{d}\Pmed_{d,d-1}=c\;\;\;\;(c\;\;\text{constant}).\ee
In particular, as $\Phi_d=\Phi_{-d}$ for any $d \in \Z$ (this is due
to the uniqueness of the invariant measure and to the symmetry of $\Pmed$),
it suffices to substitute values $d=0$ and $d=1$ in (\ref{c_costant})
to conclude that $c=0$; hence, relation (\ref{inv_rel}) holds.
\qed\end{proof}
Notice that $c=0$ corresponds to the property of time-reversibility of a Markov Chain (see Section 4.8 of \cite{ross}), hence one could
even prove it by Theorem 4.2 in \cite{ross}, after having introduced the concepts of aperiodicity and ergodicity of a Markov Chain.

From Proposition \ref{inv_prob_vector} we deduce in particular that for any $d\in\Z$, $\Phi_d>0$. Moreover, since
$\Pmed_{i-1,i}/\Pmed_{i,i-1}
< 1 $ for $i\geq 1$, $\Phi_{d}$ has a maximum at $d=0$ and it is monotone decreasing for $d>0$.

As a consequence of Proposition \ref{irr+posrec},
\begin{corollary}\label{mean1}
Let $\qmd =\Prob[\UE_k \neq
U_k|D_k=d]=\Pmed_{d,d+1}+\Pmed_{d,d-1}$, then
\bea\label{asymptotic} \lim_{K\to \infty}\Prob_b(e)=\sum_{d \in \Z}\qmd\Phi_d.\eea
\end{corollary}
\begin{proof}
Since \bea \Prob_b(e)=
\frac{1}{K}\sum_{k=0}^{K-1}\sum_{d \in \Z}\qmd
\Prob(D_k=d)=\frac{1}{K}\sum_{k=0}^{K-1}\sum_{d \in \Z}
\qmd(\Pmed^k)_{0,d}\eea the result follows from Proposition \ref{irr+posrec} and by the Lebesgue's Dominated Convergence Theorem.
Indeed,
\bea
\frac{1}{K}\sum_{k=0}^{K-1}\sum_{d \in \Z}
\qmd(\Pmed^k)_{0,d}=\sum_{d \in \Z}
\qmd\left(\frac{1}{K}\sum_{k=0}^{K-1}(\Pmed^k)_{0,d}\right)
\eea
where $\frac{1}{K}\sum_{k=0}^{K-1}(\Pmed^k)_{0,d}\leq1$.

\qed\end{proof}
This concludes the computation of the BER in case of long-time transmission, given the distribution of the input source. In the next
paragraph we study how the performance depends on the transmitted input sequence.
\subsubsection{The Conditional BER}
In the asymptotic case, the CBER converges to the same limit of the BER for \textit{almost all} the possible inputs:
\begin{theorem}\label{strong1}
Let $\pi$ be the uniform Bernoulli probability measure over $\{0,1\}^{\N}$. Then, for the One State algorithm,
\bea\lim_{K\to \infty}\Prob_b(e|\RU)=\lim_{K\to \infty}\Prob_b(e)\;\;\;\;\text{for}\;\;\pi\text{-a.e.}\;
\RU. \eea
\end{theorem}
Theorem \ref{strong1} gives a stronger result than Corollary
\ref{mean1}: the mean behavior of the One State algorithm is stated to be the behavior for each possible
input occurrence,
except for a $\pi$-negligible set. To prove Theorem \ref{strong1}, we will refer to the theory of Markov Chains in Random Environments
(see Sections \ref{par_MCRE} and \ref{par_proof1} in the Appendix).
\subsection{Theoretic Analysis of the Two States Algorithm}
Similar to the One State algorithm, the Two States procedure can be studied through the Markov Theory, which provides the instruments
to compute both BER and CBER. As shown in Section \ref{par_2states}, the Two States procedure stores, at each step, a state and its
normalized probability, this information being sufficient to individuate also the second state and  probability. Let $\XE_k$ be the
r.v. representing the stored state, $X_k$ the current correct state, $D_k=\XE_k-X_k$ and $A_k$ the r.v corresponding to the
probability of $\XE_k$: now, the stochastic process $(A_k,D_k)_{k \in \N}$ in $[0,1]\times \Z$ is a Markov Process, whose definition
(which actually extends the definition of Markov Chain from a denunerable to a continuous set) is now given.

\subsubsection{Markov Processes}\label{MP_section}
The definitions and results introduced in this Section can be retrieved in \cite{mey:93} or in the Chapter 2 of \cite{ler:03}
Consider a set $\X$ endowed with a countably generated $\sigma$-field $\FF$. A \textit{transition probability kernel} (or \textit{Markov
probability kernel}, see, e.g., \cite[Section 3.4.1]{mey:93}) on $(\X,\FF)$ is an application $P:\X\times \FF\to[0,1]$ such that\\
(i) for each $F \in \FF$, $P(\cdot, F)$ is a non-negative measurable function;\\
(ii) for each $x \in \X$, $P(x, \cdot)$ is a probability measure (p.m. for short) on $(\X,\FF)$.

Given a bounded measurable function $v$ on $(\X,\FF)$, we denote by $Pv$ the bounded measurable function on $(\X,\FF)$ defined as
\be\begin{split}
(Pv)(x)&=\int_{\X}v(y)P(x,dy).\\
   \end{split}
\ee
Further, let $\mu$ be a measure on $(\X,\FF)$: we define the measure $\mu P$
\be(\mu P)(F)=\int_{\X}P(x,F)\mu(dx)\;\;\;\;\;F \in \FF.\ee
We define the $n$-th power of the transition kernel $P$ simply putting $P^1(x,F)=P(x,F)$ and  $P^n(x,F)=\int_{\X}P(x,dy)P^{n-1}(y,F)$.
It is easy to see that $P^n(x,F)$ are transition kernels, too. Corresponding actions on bounded functions and on measures will be
respectively denoted by  $P^nv$ and $\mu P^n$.
\begin{definition}\cite[(10.1)]{mey:93}
A measure $\psi$ on $(\X, \FF)$  is said to be \emph{invariant} for the transition kernel $P$
if $\psi P=\psi$.
\end{definition}
We define a \emph{homogeneous Markov Process on space $(\X,\FF)$ with transition kernel $P$} as a sequence of $\X$-valued random
variables $(X_n)_{n \in \N}$ such that, for any $x \in \X$ and $F \in \FF$, \bea\text{Prob}(X_{n+1}\in
F|X_n=x,X_{n-1},\dots,X_0)=\text{Prob}(X_{n+1}\in F|X_n=x)=P(x,F)\eea for any $n \in \N$. The evolution of $(X_n)_{n \in \N}$ is
completely described once we fix a probability law $\mu$ of $X_0$ on $(\X,\FF)$; if $\mu$ is invariant, then the Markov Process is said
to be stationary: all the r.v.'s $X_n$ are distributed according to $\mu$.
Notice also that  for any $x \in \X$ and $F \in \FF$, $\text{Prob}(X_{m+n}\in F|X_m=x)=P^n(x,F)$ for any $m,n\in\N$.

From now onwards, we will assume that \emph{$\X$ is a locally compact separable metric space}: under this topological condition we can
easily prove the existence of an invariant measure (see \cite[Section 12.3]{mey:93}). Let $\FB(\X)$ be the Borel $\sigma$-algebra of
$\X$.
\begin{definition}\cite[Sections 6.1.1, 11.3.1]{mey:93}
Let $P$ be a transition kernel on $(\X,\FB(\X))$. If $P(\cdot,O)$ is a lower semicontinuous function for any open set $O \in \FB(\X)$,
then $P$ is said to be \emph{weak Feller}. Moreover, we say that $P$ verifies the \emph{Drift Condition} if there exist a compact set
$\CC\subset\X$, a constant $b<\infty$ and a function $V:\X\to [0,\infty]$ not always infinite such that \be\label{Drift}\Delta
V(x):=\int_{\X}P(x,dy)V(y)-V(x)\leq -1 +b\mathds 1_{\CC}(x)\ee for every  $x \in \X$. \end{definition}
\begin{proposition}\emph{\cite[Theorem 12.3.4]{mey:93}}\label{12.3.4}
If a transition kernel $P$ is weak Feller and verifies the Drift Condition, then it admits an invariant p.m..
\end{proposition}
Under some further conditions, also the uniqueness of the invariant measure can be proved.
\begin{definition}\cite[Section 4.2.1]{mey:93}
For any $B \in \mathcal{B}(\X)$, let $\tau_B=\min\{n>0: X_n \in B\}$. $(X_n)_{n \in\N}$ is said to be $\mu$\textit{-irreducible} if
there exists a measure $\mu$ on $\mathcal{B}(\X)$ such that for every $x \in \X$, $\mu(B)>0$ implies $\Prob(\tau_B<+\infty|X_0=x)>0$.
\end{definition}
A $\mu$-irreducible Markov Process whose kernel admits an invariant p.m. is said to be \emph{positive recurrent}{mey:93} and
\begin{proposition}\emph{\cite[Theorem 10.0.1, Proposition 10.1.1]{mey:93}}\label{unique_measure}
The kernel of a positive recurrent Markov Process admits a unique invariant p.m..
\end{proposition}
Furthermore,
\begin{definition}\label{def_erg}\cite[Definitions 2.2.2, 2.4.1]{ler:03}
A set $B\in \mathcal{B}(\X)$ is said to be \emph{invariant} if $P(x,B)\geq \mathds{1}_B(x)$ for every $x \in \X$.\\
A p.m. $\mu$ on $\mathcal{B}(\X)$ is said to be \emph{ergodic} if $\mu(B)=0$ or $\mu(B)=1$ for every invariant set $B \in
\mathcal{B}(\X)$.
\end{definition}
\begin{proposition}\emph{\cite[Proposition 2.4.3]{ler:03}}\label{unique_ergodic}
If a Markov Process admits a unique invariant p.m. $\mu$, then $\mu$ is ergodic.
\end{proposition}
A fundamental issue for our analysis is the Ergodic Theorem of Markov Processes, which is the transposition into stochastic terms of the
Birkhoff's Individual Ergodic Theorem (\cite[Theorem 1.14]{wal:2000}). Here we report its version under the ergodicity condition for an
invariant p.m.; for a more general treatise, see \cite{fog:69,ler:03}.
\begin{theorem}\emph{(Ergodic Theorem) \cite[Theorem 2.3.4 - Proposition 2.4.2]{ler:03}}\label{Ergodic_T} \\
Assume that a kernel $P$ on $(\X,\mathcal{B}(\X))$ admits an ergodic invariant p.m. $\mu$.
Then, for any non-negative function $v \in L_1(\X,\mathcal{B}(\X),\mu)$, \bea
\lim_{K\to\infty}\frac{1}{K}\sum_{k=0}^{K-1}(P^kv)(x)=\int_{\X}v~ d\mu\;\;\;\text{for }\; \mu\text{-a.e. } x \in \X. \eea
\end{theorem}
Finally, we report a result of direct convergence for the iterates of the kernel, in the case of no periodic behavior.
\begin{definition}\label{strape}\cite[Section 3.6]{twe:01}
A Markov Process is said to be \emph{strongly aperiodic} it there exist a set $A\subseteq \X$, a measure $\nu$ and a constant $c$ such
that $ P(x,B)\geq c\nu(B)$  for any $x \in A, B \in \FB(\X)$.
\end{definition}
Now, let $||P^n(x,\cdot)-\mu||=2\underset{B \in \FB( X) }{\sup}|P^n(x,B)-\mu(B)|$ be the total variation norm between the measures
$P^n(x,\cdot)$ and $\mu$.
\begin{proposition}\cite[Proposition 3.8]{twe:01}\label{directconv}
For a positive recurrent, aperiodic Markov Process with invariant p.m. $\mu$, $||P^n(x,\cdot)-\mu||\to 0$ as $n\to\infty$ for $\mu$-a.e.
$x\in\X$.
\end{proposition}

\subsubsection{The Mean BER}
Let $A_k$ be the r.v. representing the normalized probability of the stored state in the Two States algorithm. We observe that
$(A_k,D_k)_{k \in \N}$ is a Markov Process in  $([0,1]\times\Z,\mathcal{B}([0,1])\times \mathcal P(\Z))$ where
$\mathcal{B}([0,1])$ is the Borel $\sigma$-field on $[0,1]$ and $\mathcal P(\Z)$ is the discrete $\sigma$-field of $\Z$.
In order to completely define the process, we provide also an initial distribution $\mathcal
L\times\kappa$, $\mathcal L$ and $\kappa$ respectively
being the usual Lebesgue measure on $[0,1]$ and the counting measure on $\Z$.

The transition probability kernels will be explicitly computed in the Appendix \ref{par_trans_prob}.
\begin{proposition}\label{admits_invariant}
The kernel of $(A_k,D_k)_{k \in \N}$ admits an invariant p.m. $\widetilde{\phi}$.
\end{proposition}
\begin{proof}
We prove that the kernel of $(A_k,D_k)$ satisfies both the Weak Feller Property and the Drift Condition; the result will then follow
from Proposition \ref{12.3.4}.
First, we check the Drift Condition. By equations (\ref{sum_up1})- (\ref{sum_up3}) in the Appendix,
\be\label{Prob_d+1_d-1}
\begin{split}
P\big((\alpha,d),[0,1]\times\{d+1\}\big)&=\frac{1}{4}\text{erfc}\left(\frac{\sigma^2\log\sqrt{\frac{\alpha}{1-\alpha}}+d+1}{\sigma\sqrt{
2}}\right)\\
P\big((\alpha,d),[0,1]\times\{d-1\}\big)&=\frac{1}{2}-\frac{1}{4}\text{erfc}\left(\frac{\sigma^2\log\sqrt{\frac{\alpha}{1-\alpha}}+d}{
\sigma\sqrt{2}}\right).
\end{split}\ee
In particular, $P\big((\alpha,d),[0,1]\times\{d+1\}\big)$ and $P\big((\alpha,d),[0,1]\times\{d-1\}\big)$ have values in $[0,1/2]$ and
are monotone respectively decreasing and increasing with respect to $\alpha$. Now, let us define
 \be\label{deltad}\delta_d=\frac{1}{2(|d|+10)}\ee and
\be V(\alpha,d)=\left\{\begin{array}{ll}
d^2&~~\text{    if  } d\geq0,\alpha\geq\delta_d \text{ or if } d<0,\alpha\leq 1-\delta_d;\\
d^2+2|d|&~~\text{    otherwise.}\\
\end{array}\right. \ee
We are going to prove that  $V$ fulfills the Drift inequality for some compact $\mathrm{C}$:
\be\label{Drift2}\Delta V(\alpha,d)=\int_{[0,1]\times\Z}P\big((\alpha,d),\mathrm d(\alpha', d')\big)V(\alpha',d')-V(\alpha,d)\leq
-1+b\mathds 1_{\mathrm C}(\alpha,d)\ee for every  $(\alpha,d) \in[0,1]\times \Z$. In order to individuate $\mathrm{C}$, let us find out
the
values of $(\alpha, d)$ such that (\ref{Drift2}) holds with $\mathds 1_{\mathrm C}(\alpha,d)=0$. Recall that $P\big((\alpha,d)$,
$A\times
\{d'\}\big)>0 \Rightarrow d'\in\{d-1,d,d+1\}$ for any $\alpha \in [0,1]$, $A \in \FB([0,1])$.\\
In the next, let us use the notation $\omega=(\alpha,d)$, $\omega'=(\alpha',d')$.

If $d\geq 0$,\bea\begin{split}
&\Delta V(\omega)=\int_0^1\sum_{d'=d-1}^{d+1}P(\omega,(\mathrm d\alpha',d'))V(\omega')-V(\omega)\\
&=\sum_{d'=d-1}^{d+1}\left[\int_0^{\delta_d}P(\omega,(\mathrm d\alpha',d'))(2d'+d'^2)+\int_{\delta_d}^{1}P(\omega,(\mathrm
d\alpha',d'))d'^2\right]-V(\omega)\\
&=\sum_{d'=d-1}^{d+1}\left[\int_0^{1}P(\omega,(\mathrm d\alpha',d'))d'^2+\int_{0}^{\delta_d}P(\omega,(\mathrm
d\alpha',d'))2d'\right]-V(\omega)\\
&=\sum_{d'=d-1}^{d+1}\left[P(\omega,[0,1]\times\{d'\})d'^2+P(\omega,[0,\delta_d]\times\{d'\})2d'\right]-V(\omega)\\
&=d^2+2d[P(\omega,[0,1]\times\{d+1\})-P(\omega,([0,1]\times\{d-1\})]\\
&~~~+P(\omega,[0,1]\times\{d+1\})+P(\omega,[0,1]\times\{d-1\})+2d P(\omega,[0,\delta_d]\times\Z)\\
&~~~+2[P(\omega,[0,\delta_d]\times\{d+1\})-P(\omega,[0,\delta_d]\times\{d-1\})]
-V(\omega).\end{split}\eea
As $P(\omega,[0,1]\times\{d+1\})+P(\omega,[0,1]\times\{d-1\})\leq \frac{1}{2}$ (see equations (\ref{Prob_d+1_d-1})) and
$P(\omega,[\beta_1,\beta_2]\times\Z)\leq G(\beta_2-\beta_1)$ (see Lemma \ref{upper_bound} in the Appendix \ref{sec_ub}).
\be\label{deltaV_ineq}\begin{split}
\Delta &V(\omega)\leq
d^2+2d[P(\omega,[0,1]\times\{d+1\})-P(\omega,[0,1]\times\{d-1\})]+\frac{1}{2}\\&~~~~~~~~~~+2(d+1)G\delta_d-V(\omega)\\
&\leq d^2+2d[P(\omega,[0,1]\times\{d+1\})-P(\omega,[0,1]\times\{d-1\})]+\frac{1}{2}+G-V(\omega)
\end{split}\ee
where we exploited that $2(d+1)G\delta_d<G$ by the definition (\ref{deltad}) of $\delta_d$.

If $d<0$, by analogous computation we obtain again the inequality (\ref{deltaV_ineq}). Let us study the behavior of this bound for every
$\omega \in [0,1]\times\Z$, according to the partition of $[0,1]\times\Z$ into four subsets given by the definition of $V$.

\textbf{Subset 1}: If $d\geq 0$ and $\alpha\geq\delta_d$, $V(\omega)=d^2$ and
\bea
\begin{split}
P(\omega,[0,1]\times\{d+1\})&\leq\frac{1}{4}\text{erfc}\left(\frac{\sigma^2\log\sqrt{\frac{\delta_d}{1-\delta_d}}+d}{\sigma\sqrt{2}}
\right)\\
P(\omega,[0,1]\times\{d-1\})&\geq\frac{1}{2}-\frac{1}{4}\text{erfc}\left(\frac{\sigma^2\log\sqrt{\frac{\delta_d}{1-\delta_d}}+d}{
\sigma\sqrt{2}}\right)\\
\end{split}
\eea
hence inequality (\ref{deltaV_ineq}) becomes
\bea\begin{split} \Delta V(\omega)&\leq
G+d\left[\text{erfc}\left(\frac{\sigma^2\log\sqrt{\frac{\delta_d}{1-\delta_d}}+d}{\sigma\sqrt{2}}\right)-1\right]+\frac{1}{2}
\\&=G+d\left[\text{erfc}\left(\frac{-\frac{\sigma^2}{2}\log(2d+19)+d}{\sigma\sqrt{2}}\right)-1\right]+\frac{1}{2}.
\end{split}\eea
As $\text{erfc}(x)\in(1,2)$ when the argument $x$ is negative, then for $d$ is sufficiently large the quantity in the square bracket is
negative. Moreover, this quantity is multiplied by $d$; hence, there necessarily exists an integer $d_0^+>0$, depending on the noise
$\sigma$, such that for any $d>d_0^+$, $\Delta V(\omega)\leq-1$.\\

\textbf{Subset 2}: If $d<0$ and $\alpha\leq1-\delta_d$ ,
\bea
\begin{split}
P(\omega,[0,1]\times\{d+1\})&\geq\frac{1}{4}\text{erfc}\left(\frac{-\sigma^2\log\sqrt{\frac{\delta_d}{1-\delta_d}}+d+1}{\sigma\sqrt{2}}
\right)\\
P(\omega,[0,1]\times\{d-1\})&\leq\frac{1}{2}-\frac{1}{4}\text{erfc}\left(\frac{-\sigma^2\log\sqrt{\frac{\delta_d}{1-\delta_d}}+d+1}{
\sigma\sqrt{2}}\right)\\
\end{split}
\eea
hence inequality (\ref{deltaV_ineq}) becomes
\bea\begin{split} \Delta V(\omega)&\leq
G+d\left[\text{erfc}\left(\frac{-\sigma^2\log\sqrt{\frac{\delta_d}{1-\delta_d}}+d+1}{\sigma\sqrt{2}}\right)-1\right]+\frac{1}{2}
\\&=G+d\left[\text{erfc}\left(\frac{\frac{\sigma^2}{2}\log(-2d+19)+d+1}{\sigma\sqrt{2}}\right)-1\right]+\frac{1}{2}.
\end{split}\eea
The computation is now analogous to the previous case and we conclude that there necessarily exists an integer $d_0^-<0$, depending on
the noise, such that for any $d<d_0^-$, $\Delta V(\omega)\leq-1$.\\

\textbf{Subset 3}: If $d\geq0$ and $\alpha<\delta_d$, $V(\omega)=d^2+2d$; moreover, we have no tight bounds for
$P(\omega,[0,1]\times\{d+1\})$ and $P(\omega,[0,1]\times\{d-1\})$: we can just notice that their difference is smaller than
$\frac{1}{2}$. Substituting it in (\ref{deltaV_ineq}) we obtain
\bea \Delta V(\omega)\leq d^2 +G+\frac{1}{2}+ d-d^2-2d=G+\frac{1}{2}-d \eea
hence $\Delta V(\omega)\leq -1$ if $d>d_1=G+\frac{3}{2}$.\\

\textbf{Subset 4}: If $d<0$ and $\alpha>1-\delta_d$, $V(\omega)=d^2-2d$; as
$P(\omega,[0,1]\times\{d+1\})-P(\omega,[0,1]\times\{d-1\})\geq -\frac{1}{2}$,
\bea \Delta V(\omega)\leq G+\frac{1}{2}+d \eea
and $\Delta V\leq -1$ if $d<-d_1$.

Now, it is easy to verify that the subsets of $[0,1]\times \Z$ not yet considered form the compact set
$\big([0,\delta_d]\times\{0,\dots,d_1\}\big)\cup\big([\delta_d,1]\times\{0,\dots,d_0^+\}\big)\cup\big([0,1-\delta_d]\times\{d_0^-,\dots,
-1,0\}\big)\cup\big([1-\delta_d,1]\times\{-d^1,\dots,-1,0\}\big)$. For simplicity, we can consider the bigger compact set
$\mathbf{\mathrm{C}}=[0,1]\times\{-d_{\mathbf{\mathrm{C}}},\dots,d_{\mathbf{\mathrm{C}}}\}$, where
$d_{\mathbf{\mathrm{C}}}=\max\{d_0^+,-d_0^-,d_1\}$: now,  it is easy to check that for any $\omega \in \mathbf{\mathrm{C}}$ the Drift
Condition is satisfied whenever $b\geq G+d_{\mathbf{\mathrm{C}}}+\frac{3}{2}$.

We now check the Weak Feller Property.
Given any open interval $I\subset[0,1]$ and $d'\in\Z$,  the continuity of $P(\cdot, I\times\{d'\})$ can be easily verified by the
equations (\ref{sum_up1})-(\ref{sum_up3}) (Section \ref{par_trans_prob}): $P((\alpha,d),I\times\{d'\})$ is piecewise
defined as combination of $H$,
which is a continuous function; moreover, it is straightforward to check that the continuity  holds also at the
connection points.

 Furthermore,\\
(a) any open set on the real line (hence on $[0,1]$) is a countable union of disjoint intervals;\\
(b) if $f_N$ is a monotone increasing sequence of lower semicontinuous functions such that $f_N \uparrow f$ pointwise, then $f$ is lower
semicontinuous.

By (a), any open set $O$ in $[0,1]$ can be expressed as $O=\cup_{n=1}^{\infty}I_n$, with $I_n$ mutually disjoint open intervals in
$[0,1]$.  Moreover, $f_N(\omega)=P(\omega, (\cup_{n=1}^{N}I_n)\times\{d'\})\leq 1$ fulfills the hypotheses of statement (b), hence its
pointwise limit $f(\omega)=P(\omega, (\cup_{i=1}^{\infty}I_i)\times\{d'\})=P(\omega, O\times\{d'\})$ is lower semicontinuous. As any
open set of the product topology can be expressed as $\cup_{n \in \Z}(O_n\times\{n\})$, $O_n$ open in $[0,1]$, the lower semicontinuity
is extended to all the open sets.
\qed\end{proof}
Given the existence of an invariant p.m., we now evaluate the BER by means of the Ergodic Theorem \ref{Ergodic_T}.  The BER is given by
\bea\begin{split}P_b(e)
&=\frac{1}{K}\sum_{k=0}^{K-1}P(\UE_k \neq
U_k)=\frac{1}{K}\sum_{k=0}^{K-1}\int_0^1\sum_{d \in \Z}P(\UE_k \neq
U_k,A_k=\alpha,D_k=d)\mathrm d\alpha\\
&=\frac{1}{K}\sum_{k=0}^{K-1}\int_0^1\sum_{d \in \Z}P(\UE_k \neq
U_k|A_k=\alpha,D_k=d)P^k\big((1,0);(\mathrm d\alpha,d)\big).
\end{split}\eea
the initial state $(1,0)$ being discussed in the Remark \ref{initialstate}.
Let $\qmad(\alpha,d)=P(\UE_k \neq
U_k|\alpha_k=\alpha,D_k=d)$ (notice that  $\qmad(\alpha,d)$ actually does not depend on $k$) so that
$P_b(e)=\frac{1}{K}\sum_{k=0}^{K-1}(P^k \qmad)(1,0)$.
Then,
\begin{corollary}\label{mean2}
Given the invariant p.m. $\widetilde{\phi}$,
\bea\begin{split}\lim_{K\to\infty}P_b(e)=
\int_{[0,1]\times\Z}\qmad~\mathrm d\widetilde{\phi}.\end{split}\eea
\end{corollary}
\begin{proof}
$(A_k,D_k)_{k \in \N}$ is $(\mathcal
L\times\kappa)$-irreducible (the proof of this fact requires some technical computation and is postponed in the Appendix
\ref{par_irred}), then $\widetilde{\phi}$ is unique and ergodic by Propositions \ref{unique_measure} and \ref{unique_ergodic}.
Therefore, by the Ergodic Theorem \ref{Ergodic_T}, 
\bea\lim_{K\to\infty} \frac{1}{K}\sum_{k=0}^{K-1}(P^k \qmad)(\alpha,d)=\int_{[0,1]\times\Z}\qmad~\mathrm
d\widetilde{\phi}~~~\widetilde{\phi}\text{-a.e. }(\alpha,d).\eea
This result cannot be immediately applied to evaluate the BER since the convergence is not assured for \emph{all} the initial states. In
particular, let call $N\subset [0,1]\times \Z$  the negligible set for which there is no convergence and let $N_0=\{\alpha \in [0,1]:
(\alpha,0)\in N\}$. Now, recalling the Remark \ref{extreme},
\bea\begin{split}P_b(e)
&=\frac{1}{K}\qmad(1,0)+   \frac{1}{K}\sum_{k=1}^{K-1}\int_{\alpha_1\in[0,1]} \sum_{d_1\in\Z}P((1,0),(\mathrm d \alpha_1,d_1)) (P^{k-1}
\qmad)(\alpha_1,d_1)\\
&=\frac{1}{K}\qmad(1,0)+   \frac{1}{K}\sum_{k=1}^{K-1}\int_{\alpha_1\in[0,1]}P((1,0),(\mathrm d \alpha_1,0)) (P^{k-1}
\qmad)(\alpha_1,0).
\end{split}\eea
By the Lebesgue's Dominated Convergence Theorem,
\bea\lim_{K\to\infty}P_b(e)
=\int_{\alpha_1\in[0,1]}P((1,0),(\mathrm d \alpha_1,0))\lim_{K\to\infty}\frac{1}{K}\sum_{k=1}^{K-1} (P^{k-1} \qmad)(\alpha_1,0).\eea
Notice that $\mathcal{L}(N_0)=0$, otherwise $\widetilde{\phi}(N_0\times\{0\})=\int_{[0,1]\times
\Z}P(\omega,N_0\times\{0\})\widetilde{\phi}(\mathrm d \omega)>C_{\ep,0}\mathcal{L}(N_0)>0$ by Proposition\ref{lower_bound}. By
Proposition \ref{upper_bound}, this implies that $P((1,0),N_0\times\{0\})=0.$
Finally, 
\bea\begin{split}\lim_{K\to\infty}P_b(e)
&=\int_{\alpha_1\in[0,1]\setminus N_0}P((1,0),(\mathrm d \alpha_1,0))\lim_{K\to\infty}\frac{1}{K}\sum_{k=1}^{K-1} (P^{k-1}
\qmad)(\alpha_1,0)\\
&=\int_{\alpha_1\in[0,1]\setminus N_0}P((1,0),(\mathrm d \alpha_1,0))\int_{[0,1]\times\Z} \qmad~\mathrm d
\widetilde\phi=\int_{[0,1]\times\Z} \qmad~\mathrm d \widetilde\phi\\
\end{split}\eea
as $(\alpha_1,0)\notin N$.
\qed\end{proof}
The function $\qmad(\alpha, d)$ is explicitly computed in the Appendix \ref{par_qmad}.
\subsubsection{The Conditional BER}
The CBER for the Two States algorithm can be derived just as we computed it for the One State case, in fact it holds the following
\begin{theorem}\label{strong2}
Let $\pi$ be the uniform Bernoulli probability measure over $\{0,1\}^{\N}$. Then, for the Two States algorithm,
\bea\lim_{K\to \infty}P_b(e|\RU)=\lim_{K\to \infty}P_b(e)\;\;\;\;\text{for}\;\;\pi\text{-a.e.}\;
\RU. \eea
\end{theorem}
We refer the reader to the Appendix \ref{par_proof2} for the proof.
\subsection{Direct Convergence to $\widetilde{\phi}$}
The explicit construction of an invariant p.m. is an intricate issue in the not countable framework. When ergodic results are available,
one can approximate it by several procedures (see, e.g, \cite[Chapter 12]{ler:03}).  In our framework, we can obtain an approximation by
Proposition \ref{directconv}, which states the direct convergence of the iterates $P^n(\cdot,\cdot)$ to the invariant p.m.. Before
illustrating that, let us prove that the hypotheses of  Proposition \ref{directconv} hold.
\begin{proposition}\label{aperiod}
The Markov Process $(A_k,D_k)_{k\in\N}$ is strongly aperiodic.
\end{proposition}
\begin{proof}
Let us consider the probability measure $\mathcal{L}\times \delta_{\bar{d}}$ on $([0,1]\times\Z,\FB([0,1])\times \mathcal P(\Z))$, where
$\mathcal{L}$ is the Lebesgue measure and $\delta_{\bar{d}}(d)=1$ if $d=\bar{d}$, 0 otherwise. By Proposition \ref{lower_bound},
$P((\alpha,d), M\times\{d\})>\frac{1}{2}C_{\ep,d}\mathcal{L}(M)$, $C_{\ep,d}>0$. Then, considering the Definition \ref{strape} with
$\nu=\mathcal{L}\times \delta_{\bar{d}}$, $c=\frac{1}{2}C_{\ep,d}$ and $A=[0,1]\times \{\bar{d}\}$, the proposition is
proved.\qed\end{proof} 
This result along with Proposition \ref{directconv} yields:
\begin{corollary}[Direct Convergence]\label{phitilde_directconv}
$||P^n((\alpha,d),\cdot)-\widetilde{\phi}||\to 0$ as $n\to\infty$ for $\phi$-a.e. $(\alpha,d)\in [0,1]\times\Z$. 
\end{corollary}

\subsubsection{Analytic vs Simulations' outcomes}
\begin{figure}[h]
  \begin{center}
  \includegraphics[width=9cm, viewport=60 70 770 550]{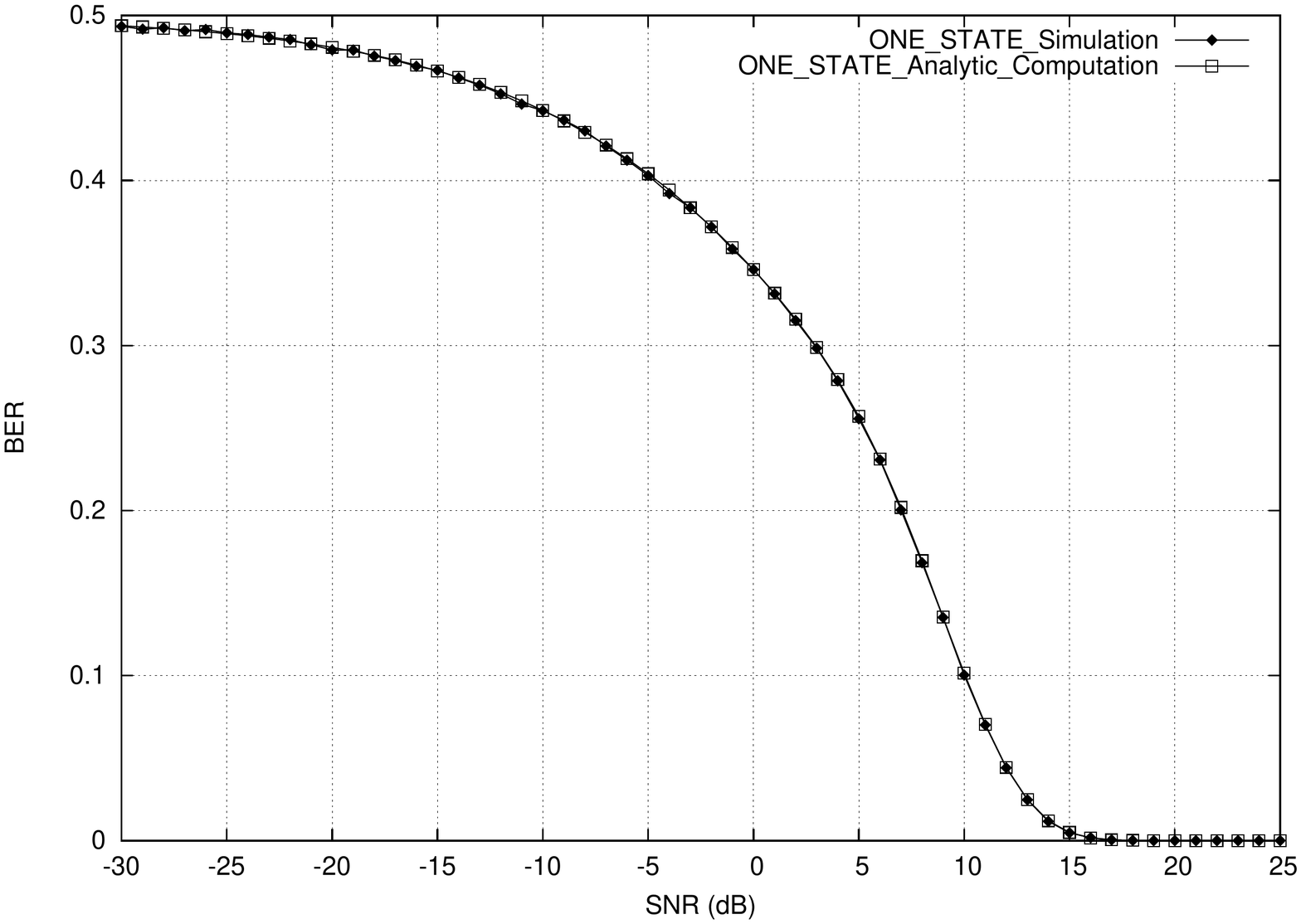}
  \caption{One State: analytic computation vs simulation.}\label{1state_figure}
  \end{center}
  \begin{center}
  \includegraphics[width=9cm, viewport=60 70 770 550]{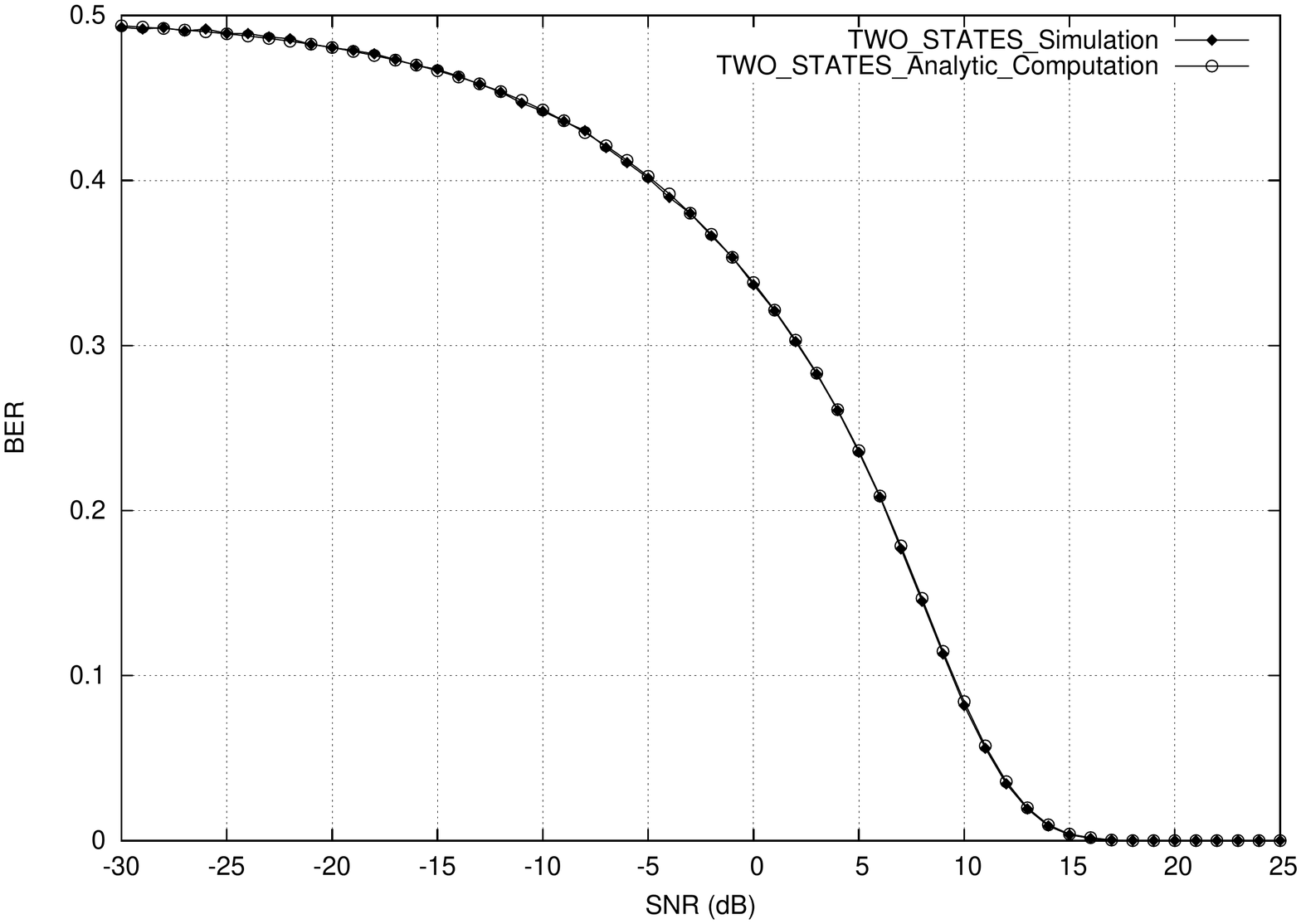}
  \caption{Two States: analytic computation vs simulation.}\label{2states_figure}
  \end{center}
\end{figure}
To conclude our analysis of One State and Two States algorithms, we compare the simulations' outcomes with the theoretic results: we
expect the BER's obtained by the simulations of sufficiently long transmissions to be consistent to the analytic computations.

By Corollaries \ref{mean1} and \ref{mean2}, the BER's can be computed once we know the corresponding invariant distributions.
While for the One State algorithm the invariant measure is explicitly given by (\ref{inv_rel}), for the Two States algorithm
we have approximated it using the Corollary \ref{phitilde_directconv}. In particular, we have discretized the kernel $P$ into a matrix,
afterwards we have computed the iterates $P^n$ for a sufficiently large $n$, so that to obtain an equilibrium condition, that is, a
matrix whose rows are all equal up to numerical roundoff . At this point, any row of the matrix is a discretized, approximated version
of the invariant p.m.


In Figures \ref{1state_figure} and \ref{2states_figure}, we compare analytic and simulations' outcomes: as expected, they do not present
substantial differences.
\section{Appendix}
\subsection{Markov Chains in Random Environments}\label{par_MCRE}
Consider a countable set $\Theta$ and a family of transition probability kernels $\{P_{\theta}, \theta \in \Theta \}$ on a space
$(\X,\FF)$. Given a $\sigma$-field $\FB$ of $\Theta$, let $(\theta_n)_{n \in \N}$ and $(X_k)_{k \in \N}$ respectively be sequences of
$\Theta$-valued and $\X$-valued r.v's. $P_{\theta_k}(X_k,F)$ can now be interpreted as the transition probability of $X_k$ to set $F$
depending on the r.v $\theta_k$, which represents to so-called \emph{random environment}.\\
We say that $(X_k)_{k \in \N}$ with $(\theta_n)_{n \in \Z}$ is a Markov Chain in Random Environment (or MCRE) if
\be\begin{split} P(&X_{k+1} \in F|X_{k},\dots,X_0,(\theta_n)_{n\in\Z})=P_{\theta_k}(X_k,F)\;\;\text{ a.s.}\\&\text{for all }F \in \FF
\text{ and } k=0,1,\dots \end{split}\ee
Let us define $\Theta^{\N}=\prod_{0}^{+\infty}\Theta$ and $\FB^{\N}=\prod_{0}^{+\infty}\FB$.
An important feature of a MCRE is that we can always associate to it a classical Markov Process. In fact,
given any $x \in \X$ and $\underline \theta=(\theta_0,\theta_1,\dots) \in \Theta^{\N}$ and denoting by $T$ the left sequence shift on
$\Theta^{\N}$ (that is, $T\underline \theta=\widetilde{\underline{\theta}}$ with $\widetilde{\underline \theta}_n=\underline
\theta_{n+1}$ for any $n \in \N$), we can introduce the following transition probability kernel on $(\X\times
\Theta^{\N},\FF\times\FB^{\N})$:
\be P\big((x,\underline \theta),F\times B\big)=P_{\theta_0}(x,F)\mathds 1_B(T\underline \theta) \ee
which determines a Markov Process $\big(X_{k},T^k(\theta_{n})_{n \in \N}\big)_{k \in \N}$ on $(\X\times\Theta^{\N},\FF\times\FB^{\N})$.
From now onwards, we will refer to it as to the \textit{Extended Markov Process}, EMP for short.
\begin{remark}: As noted in the Section 1 of \cite{cog:84}, if the random environments  $\theta_n$'s are independent,
then $(X_k)_{k \in \N}$ is a Markov Process with transition probability kernel 
$P(x,F)=\E_{\underline \theta \in \Theta^n}\left[P_{\theta_0}(x,F)\right]$. In
other terms, $(X_k)_{k \in \N}$ is the Markov Process moving in the \emph{average environment}.
\end{remark}
In this framework, we prove the following
\begin{proposition}\label{invariant_MCRE}
Let $(X_k)_{k \in \N}$ with $(\theta_n)_{n \in \N}$ be a MCRE on $\X\times \Theta^{\N}$. Suppose that the random environments
$\theta_n$'s are independent, identically distributed with distribution $\pi_0$ on $(\Theta,\FB)$ and that the kernel of the Markov
Process $(X_k)_{k \in \N}$  admits an invariant p.m. $\phi$; given the distribution $\pi=\times_{n=0}^{\infty}\pi_0$ over
$(\Theta^{\N},\FB^{\N})$,
\be\label{prodmeas}\psi=\phi\times\pi\ee is an invariant p.m. for the EMP $\big(X_k, T^k(\theta_n)_{n \in \N}\big)_{k \in \N}$ over
$(\X\times
\Theta^{\N},\FF\times \FB^{\N})$.
\end{proposition}
\begin{proof}
Let $\omega=(x,\underline \theta) \in \X\times \Theta^{\N}$. $\psi$ is an invariant for $(X_k, \theta_k)_{k \in \N}$ if \bea
\int_{\X\times \Theta^{\N}}P(\omega,F\times B)\psi(\mathrm d\omega)=\psi(F \times B) \eea  for any $F\times B$ such that $F \in \FF$, $B
\in \FB^{\N}$. Now,
\bea\begin{split}
\int_{\X\times \Theta^{\N}}P(\omega,F\times B)\psi(\mathrm
d\omega)&=\int_{\X}\int_{\Theta^{\N}}P_{\theta_0}(x,F)\mathds{1}_B(\theta_1,\theta_2,\dots)\pi(\mathrm d\underline \theta)\phi(\mathrm d
x)\\
&=\pi(B)\int_{\X}\sum_{\theta_0 \in \Theta}P_{\theta_0}(x,F)\pi_0(\theta_0)\phi(\mathrm d x)\\
&=\pi(B)\int_{\X}P(x,F)\phi(\mathrm d x)=\pi(B)\phi(F)=\psi(F\times B)
\end{split}\eea
where we have exploited the fact that $\phi$ is invariant.
\qed\end{proof}
This Proposition is a partial extension of the Theorem 5 in \cite{naw:81}, which states the same result in the case of denumerable state
space $\X$ and attests also the inverse implication (that is, all the invariant p.m.'s are  product measures of kind (\ref{prodmeas})
still in the denumerable framework.

For a more detailed treatise on MCRE's, we refer the reader to \cite{cog:84,cog:86,naw:81,naw:82}.
\subsection{Proof of Theorem \ref{strong1}}\label{par_proof1}
From equation \ref{D_k_rec}, $(D_k)_{k \in \N}$ with $(U_k)_{k \in \N}$ turns out to be a countable
MCRE. This is the right way to look at $(D_k)_{k \in \N}$ if we want to
understand its behavior with respect to typical instances of the
input $\RU=(U_0,U_1,\dots)$. For any $x,y \in \Z$, we have
\bea P(D_{k+1}=y|D_k=x,D_{k-1},\dots,D_0; \RU)=\mathbf{P}_{x,y}(U_k).\eea
Consider the space $(\Z\times
\{0,1\}^\N,\mathcal{P}(\Z)\times\prod_0^{\infty}\mathcal{P}(\{0,1\}))$ endowed with the initial distribution
$\kappa\times\pi$, where $\kappa$ is the counting measure on
$\Z$ and $\pi$ is the usual uniform Bernoulli measure on
$\{0,1\}^\N$. Given $x,y\in\Z$, $\su=(u_0,u_1,\dots)\in\{0,1\}^{\N}$ and $B \in
\prod_0^{\infty}\mathcal{P}(\{0,1\})$, the EMP is defined by the transition probability kernel \be P\big((x,\su);\{y\}\times
B\big)=\mathbf P_{x,y}(u_0)\mathds
1_B(T\su).\ee
By Proposition \ref{invariant_MCRE}, an invariant
probability measure exists for our EMP and we explicitly compute it: in fact, let $\phi$ be a p.m. on $(\Z,\mathcal{P}(\Z))$ given
by $\phi(\{d\})=\Phi_d$, $\Phi_d$ being the invariant probability vector defined in the Proposition \ref{inv_prob_vector}, for any
integer $d$. Then,
$\psi=\phi\times\pi$ is an invariant p.m. for the
EMP.\\
We can verify that $\psi$ is ergodic by the following criterion (see Chapter 3 of \cite{cog:86}). Let $\P(U_0,\dots U_{n-1})$ the
transition matrix whose entries are  \begin{equation}\P_{x,y}(U_0,\dots U_{n-1})=\Prob(D_{n}=y|D_0=x,U_0,\dots U_{n-1}).\end{equation}

If for each $x,y\in \Z$ and $\pi$-a.e. $\RU$ there exist $n=n(x,y,\RU)\in\N$
and $z=z(x,y,\RU,n)\in \Z$ such that $
\P_{x,z}(U_0,\dots,U_{n-1})\P_{y,z}(U_0,\dots U_{n-1})>0$, then
$\psi$ is ergodic. In our context it is easy to check that given any couple of starting states $x$ and
$y$, after $n>|x-y|$ steps we have a non-null
probability of having joined a common state $z$.\\Define $\qd(U_k)=P[\UE_k \neq U_k|D_k=d,U_k]=\P_{d,d+1}(U_k)+\P_{d,d-1}(U_k)$ ($\qmd$
is actually the mean of $\qd$).
For any $K \in \N$ and given $D_0=0$, the CBER  can be expressed as
follows: \be\begin{split} &P_b(e|\RU)=
\frac{1}{K}\sum_{k=0}^{K-1}\sum_{d\in\Z} \qd(U_k)\P_{0,d}(U_0,U_1,\dots U_{k-1})\\
    \end{split}\ee
Notice that, since the $U_k$'s $k\in\N$ are independent, $\P(U_0,U_1,\dots,U_{k-1})=\P(U_0)\P(U_1)\cdots\P(U_{k-1})$.

Consider $\omega=(x,\RU)$ and the function $g(\omega)=\qx(U_0)$: we have that  \bea\sum_{d \in \Z}\qd(U_k)\mathbf P_{x,d}(U_0,\dots
U_{k-1})=P^k g(x,\RU)\eea and notice that \be\label{pbeu} P_b(e|\RU)=\frac{1}{K}\sum_{k=0}^{K-1}P^k g(0,\RU).\ee
Now, by the Ergodic Theorem \ref{Ergodic_T}: \be\label{byERG}
\lim_{K\to\infty}\frac{1}{K}\sum_{k=0}^{K-1}P^k
g(\omega)=\int_{\Z\times\{0,1\}^{\N}}g(\omega)\psi(\mathrm d\omega)\;\;\;\text{for}\;\;\psi\text{-a.e. } \omega.\ee
Notice that, as pointed out after Proposition \ref{inv_prob_vector}, $\phi(\{d\})>0$ for any $d\in\Z$; then, a set
$\{d\}\times B$, $d\in\Z$, $B\subset \{0,1\}^{\N}$, is $\psi$-negligible if  and only if $\pi(B)=0$. 
Hence, in (\ref{byERG}), ``$\psi\text{-a.e. } \omega$'' is equivalent to ``for any $d\in\Z$ and $\pi$-a.e. $\RU$''.  

This, along with  the equality (\ref{pbeu}), implies that 
\be \lim_{K\to\infty} P_b(e|\RU)=\int_{\Z\times\{0,1\}^{\N}}g(\omega)\psi(\mathrm d\omega)\;\;\;\text{for}\;\;\pi\text{-a.e. }
\RU.\ee
Finally, recalling that
$\psi=\phi\times \pi$, \bea\int_{\Z\times\{0,1\}^{\N}}g(\omega)\psi(\mathrm d\omega)=\sum_{d \in
\Z}\sum_{U_0=0,1}\qd(U_0)\pi(U_0)\Phi_d=\sum_{d \in \Z}\qmd\Phi_d.\eea
\subsection{Two States Algorithm: Computation of the Transition Probabilities}\label{par_trans_prob}
In the next pages, we compute the probability of moving from a state $(\alpha,d)\in [0,1]\times \Z$ to a set of type
$(0,\beta)\times\{d'\}$, $\beta \in (0,1], d' \in \Z$, for the Markov Process $(A_k,D_k)_{k \in \N}$ defined in Section 3.2. Let
$P_{u}\big((\alpha,d),(0,\beta)\times\{d'\}\big)$ be the transition probability given the transmitted bit $u$:
$ P\big((\alpha,d),(0,\beta)\times\{d'\}\big)
=\frac{1}{2}P_{0}\big((\alpha,d),(0,\beta)\times\{d'\}\big)+\frac{1}{2}P_{1}\big((\alpha,d),(0,\beta)\times\{d'\}\big)$
are null if $d'\notin \{d-1,d,d+1\}$, if $d'=d+1$ and $u=1$ or if $d'=d-1$ and $u=0$; we now compute the non-null instances. Given
$(\alpha,d)\in (0,1)\times \Z$ and $x\in\{\alpha,(1-\alpha)^{-1},1\}$, $y \in \{d-1,d,d+1\}$, $z \in (0,1)$, we define:
\be\label{defchH}\begin{split}
&c_{\alpha}=\sqrt{\frac{\exp(1/\sigma^2)}{\alpha(1-\alpha)}}\\
&h_{x,y}(z)=\frac{\sigma^2\log\left(x\frac{1-z}{z}\right)+y+\frac{1}{2}}{\sigma\sqrt{2}}\\
&H_{x,y}(z)=\frac{1}{2}\text{erfc}\left(h_{x,y}(z)\right).
\end{split}
\ee
Notice that these quantities depend on the noise variance $\sigma^2$, even if the notation does not emphasize that. Remind also
Definition (\ref{zeta_i}).\\
\textbf{Case 1}: $d'=d,u=0$.
\be\label{d0}\begin{split}
&P_{0}\big((\alpha,d),(0,\beta)\times\{d\}\big)=\text{Prob}(\zeta_3\leq\zeta_1\leq\beta (\zeta_1+\zeta_2)|A_k=\alpha,D_{k}=d,U_k=0)\\
&=\left\{\begin{array}{lr}
0 &\text{ if }\alpha=0 \text{ or if } \alpha \in (0,1) \text{ and } \beta\leq\frac{1}{1+c_{\alpha}}\\
H_{\alpha,d}\left(\beta\right)-
H_{\alpha,d}\left(\frac{1}{1+c_{\alpha}}\right)&\text{ if } \alpha \in (0,1) \text{ and }\beta>\frac{1}{1+c_{\alpha}}\\
H_{1,d}\left(\beta\right)& \text{ if } \alpha=1.\\
\end{array}
\right.\\
\end{split}\ee
\textbf{Case 2}: $d'=d,u=1$.
\be\label{d1}\begin{split}
&P_{1}\big((\alpha,d),(0,\beta)\times\{d\}\big)=\\&~~~~~~~~~~~~~~~~~~=\text{Prob}\big((\zeta_3\geq\zeta_1)\cap
(\beta\zeta_3\geq(1-\beta)\zeta_2)|A_k=\alpha,D_{k}=d,U_k=1\big)\\
&=\left\{\begin{array}{lr}
H_{\frac{1}{1-\alpha},d}\left(\beta\right)&\text{ if } \alpha=0 \text{ or if } \alpha \in (0,1) \text{ and }
\beta\leq\frac{c_{\alpha}}{1+c_{\alpha}}\\
H_{\frac{1}{1-\alpha},d}\left(\frac{c_{\alpha}}{1+c_{\alpha}}\right)&\text{ if } \alpha \in (0,1) \text{ and }
\beta>\frac{c_{\alpha}}{1+c_{\alpha}}\\
0& \text{ if } \alpha=1.
\end{array}
\right.\\
\end{split}\ee
\textbf{Case 3}: $d'=d+1,u=0$.
\be\label{dp0}\begin{split}
&P_{0}\big((\alpha,d),(0,\beta)\times\{d+1\}\big)=\\&~~~~~~~~~~~~~~~~~~=\text{Prob}\big((\zeta_3\geq\zeta_1)\cap
(\beta\zeta_3\geq(1-\beta)\zeta_2)|A_k=\alpha,D_{k}=d,U_k=0\big)\\
&=\left\{\begin{array}{lr}
H_{\frac{1}{1-\alpha},d+1}\left(\beta\right)
&\text{ if }  \alpha=0 \text{ or if }\alpha \in (0,1) \text{ and } \beta\leq\frac{c_{\alpha}}{1+c_{\alpha}}\\
H_{\frac{1}{1-\alpha},d+1}\left(\frac{c_{\alpha}}{1+c_{\alpha}}\right) &\text{ if } \alpha \in (0,1) \text{ and }
\beta>\frac{c_{\alpha}}{1+c_{\alpha}}\\
0& \text{ if } \alpha=1.
\end{array}
\right.\\
\end{split}\ee
\textbf{Case 4}: $d'=d-1, u=1$.
\be\label{dm1}\begin{split}
&P_{1}\big((\alpha,d),(0,\beta)\times\{d-1\}\big)=\\&~~~~~~~~~~~~~~~~~~~~=\text{Prob}(\zeta_3\leq\zeta_1\leq\beta
(\zeta_1+\zeta_2)|A_k=\alpha,D_{k}=d,U_k=1)\\
&=\left\{\begin{array}{lr}
0 &\text{ if }\alpha=0 \text{ or if } \alpha\in(0,1)\text{ and }\beta\leq\frac{1}{1+c_{\alpha}}\\
H_{\alpha,d-1}\left(\beta\right)-
H_{\alpha,d-1}\left(\frac{1}{1+c_{\alpha}}\right)&\text{ if } \alpha \in (0,1) \text{ and }\beta>\frac{1}{1+c_{\alpha}}\\
H_{1,d-1}\left(\beta\right)& \text{ if } \alpha=1.\\
\end{array}\right.\\
\end{split}\ee
\begin{remark}\label{calphabounds}: As $c_{\alpha}> 2$,
$\frac{1}{1+c_{\alpha}}<\frac{1}{3}<\frac{2}{3}<\frac{c_{\alpha}}{1+c_{\alpha}}$.
\end{remark}

 Summing up:\be\label{sum_up1}\begin{split}& P\big((\alpha,d), (0,\beta)\times \{d\}\big)=\\&
\frac{1}{2}\left\{\begin{array}{l}
H_{1,d}\left(\beta\right)~~~~~~~~~~~~~~~~~~~~~~~~~~~~~~~~~~~~\text{ if }\alpha=0 \text{ or if } \alpha=1\\
H_{\frac{1}{1-\alpha},d}\left(\beta\right)~~~~~~~~~~~~~~~~~~~~~~~~\text{ if } \alpha \in (0,1) \text{ and }
\beta\leq\frac{1}{1+c_{\alpha}}\\
H_{\alpha,d}\left(\beta\right)-H_{\alpha,d}\left(\frac{1}{1+c_{\alpha}}\right)+H_{\frac{1}{1-\alpha},d}\left(\beta\right)\\
~~~~~~~~~~~~~~~~~~~~~~~~~~~~\text{ if } \alpha\in(0,1) \text{ and } \frac{1}{1+c_{\alpha}}<\beta\leq\frac{c_{\alpha}}{1+c_{\alpha}}\\
H_{\alpha,d}\left(\beta\right)-H_{\alpha,d}\left(\frac{1}{1+c_{\alpha}}\right)+H_{\frac{1}{1-\alpha},d}\left(\frac{c_{\alpha}}{1+c_{
\alpha}}\right)\\
~~~~~~~~~~~~~~~~~~~~~~~~~~~~~~~~~~~~~~~\text{ if } \alpha\in(0,1) \text{ and } \beta>\frac{c_{\alpha}}{1+c_{\alpha}}\\
\end{array}\right.\end{split}\ee
\be\label{sum_up2}\begin{split}
& P\big((\alpha,d), (0,\beta)\times \{d+1\}\big)=\\
&\frac{1}{2}\left\{\begin{array}{lr}
H_{\frac{1}{1-\alpha},d+1}\left(\beta\right)
&\text{ if }  \alpha=0 \text{ or if }\alpha \in (0,1) \text{ and } \beta\leq\frac{c_{\alpha}}{1+c_{\alpha}}\\
H_{\frac{1}{1-\alpha},d+1}\left(\frac{c_{\alpha}}{1+c_{\alpha}}\right) &\text{ if } \alpha \in (0,1) \text{ and }
\beta>\frac{c_{\alpha}}{1+c_{\alpha}}\\
0& \text{ if } \alpha=1
\end{array}\right.
\end{split}\ee
\be\label{sum_up3}\begin{split}
& P\big((\alpha,d), (0,\beta)\times \{d-1\}\big)=\\&\frac{1}{2}\left\{\begin{array}{l}
0 ~~~~~~~~~~~~~~~~~~~~~~~\text{ if }\alpha=0 \text{ or if } \alpha\in(0,1) \text{ and } \beta\leq\frac{1}{1+c_{\alpha}}\\
H_{\alpha,d-1}\left(\beta\right)-
H_{\alpha,d-1}\left(\frac{1}{1+c_{\alpha}}\right)~\text{ if } \alpha \in (0,1) \text{ and }\beta>\frac{1}{1+c_{\alpha}}\\
H_{1,d-1}\left(\beta\right)~~~~~~~~~~~~~~~~~~~~~~~~~~~~~~~~~~~~~~~~~~~~~~~~~ \text{ if } \alpha=1.\\
\end{array}\right.\\ \end{split} \ee
\subsection{Two States Algorithm: Computation of $\qmad(\alpha,d)$}\label{par_qmad}
The function  $\qmad$ on $[0,1]\times \Z$ defined in the Corollary \ref{mean2} is given by $\qmad(\alpha,d)
=\frac{1}{2}\Prob(\UE_k =1|
U_k=0,A_k=\alpha,D_k=d)+\frac{1}{2}\Prob(\UE_k =0|
U_k=1,A_k=\alpha,D_k=d)$.
Note that
\bea\begin{split} &\Prob(\UE_k
=1| U_k=0,A_k=\alpha,D_k=d)=\\
&=\text{Prob}\big(\alpha f_{(Y_{k+1}|X_{k+1})}(y_{k+1}|\xe_k+1)+(1-\alpha)f_{(Y_{k+1}|X_{k+1})}(y_{k+1}|\xe_k+2)\\&>\alpha
f_{(Y_{k+1}|X_{k+1})}(y_{k+1}|\xe_k)+(1-\alpha)f_{(Y_{k+1}|X_{k+1})}(y_{k+1}|\xe_k+1)\big)\\
&=\frac{1}{2}\text{erfc}\left(\frac{\sigma^2\log z_1+d+\frac{1}{2}}{\sqrt{2}\sigma}\right)\\
\end{split}\eea
where $z_1$ is the positive solution of the equation
$(1-\alpha)e^{-\frac{1}{\sigma^2}}z^2+(2\alpha-1)z-\alpha=0$.
Similarly, \bea \Prob(\UE_k =0|
U_k=1,A_k=\alpha,D_k=d)=1-\frac{1}{2}\text{erfc}\left(\frac{\sigma^2\log
z_1+d-\frac{1}{2}}{\sqrt{2}\sigma}\right)\eea
hence \bea\qmad(\alpha,d)=\frac{1}{2}\left[\frac{1}{2}\text{erfc}\left(\frac{\sigma^2\log
z_1+d+\frac{1}{2}}{\sqrt{2}\sigma}\right)+1-\frac{1}{2}\text{erfc}\left(\frac{\sigma^2\log
z_1+d-\frac{1}{2}}{\sqrt{2}\sigma}\right)\right].\eea
Naturally, if $\alpha=1$, then $\qmad
(\alpha,d)=\frac{1}{2}\left[\frac{1}{2}\text{erfc}\left(\frac{d+\frac{1}{2}}{\sqrt{2}\sigma}\right)+1-\frac{1}{2}\text{erfc}\left(\frac{
d-\frac{1}{2}}{\sqrt{2}\sigma}\right)\right]=\qmd$ and we recast into the One State case.

\subsection{Two States Algorithm: Proof of the $(\mathcal L \times \kappa)$-irreducibility of $(A_k,D_k)_{k \in \N}$}\label{par_irred}
In this paragraph, we complete the proof of the Corollary \ref{mean2} showing the $(\mathcal L \times \kappa)$-irreducibility of
$(A_k,D_k)_{k \in \N}$ in the space $([0,1]\times\Z,\FB([0,1])\times \mathcal P(\Z))$. For this purpose, we first  prove that any
non-negligible Borel subset of kind $M\times \{d'\}\subset [0,1]\times\Z$ is achievable with positive probability from any $(\alpha,d)$,
in one or two steps, if $d'\in\{d-1,d,d+1\}$ and $M$ is sufficiently far from the extreme points of $[0,1]$:
\begin{lemma}\label{lontano_dai_bordi}
For any $\varepsilon>0$, $d\in \Z$, there exists a constant $C_{\ep, d}>0$ such that the following inequalities hold  for every
$(\alpha,d) \in [0,1]\times\Z$ and $M\in \FB([\ep,1-\ep])$,:
\bea\begin{split} &P\big((\alpha,d),M\times\{d\}\big)\geq C_{\ep,d}\mathcal{L}(M)\\
&P^2\big((\alpha,d),M\times\{d+1\}\big)\geq C_{\ep,d}\mathcal{L}(M)\\
&P^2\big((\alpha,d),M\times\{d-1\}\big)\geq C_{\ep,d}\mathcal{L}(M)\end{split}\eea where $\mathcal{L}$ is the Lebesgue measure.
\end{lemma}
\begin{proof}
First, we prove the lemma on the open intervals $(\beta_1,\beta_2)\subset [\varepsilon,1-\varepsilon]$. For shortness of notation, let
$\bar{\alpha}=\frac{1}{1-\alpha}$.

Consider the first inequality. On the basis of the equations (\ref{sum_up1}) and Remark \ref{calphabounds}, the following cases may
occur:
\begin{enumerate}
 \item If $\alpha=0$, $(\beta_1,\beta_2)\in[\ep,1-\ep]$ or if $\alpha \in (0,1)$, $(\beta_1,\beta_2)\subset
[\ep,\frac{1}{1+c_{\alpha}}\frac{1}{3}]\subseteq[\ep,\frac{1}{3}]$:
\bea\begin{split}
P\big((\alpha,d),&(\beta_1,\beta_2)\times\{d\}\big)=\frac{1}{2}H_{\bar{\alpha},d}\left(\beta_2\right)-\frac{1}{2}H_{\bar{\alpha},d}
\left(\beta_1\right)\\
&=\frac{1}{2\sqrt{\pi}}\int_{h_{\bar{\alpha},d}(\beta_2)}^{h_{\bar{\alpha},d}(\beta_1)}e^{-t^2}\mathrm d t\\
&=-\frac{1}{2\sqrt{\pi}}\int_{\beta_1}^{\beta_2}e^{-h_{\bar{\alpha},d}^2(z)}\frac{\partial }{\partial z}h_{\bar{\alpha},d}(z)\mathrm d
z\\
&\geq\frac{1}{2\sqrt{\pi}}(\beta_2-\beta_1)\min_{z \in (\beta_1, \beta_2)}\left( -e^{-h_{\bar{\alpha},d}^2(z)}\frac{\partial }{\partial
z} h_{\bar{\alpha},d}(z)\right)\\
&\geq\frac{1}{2\sqrt{\pi}}(\beta_2-\beta_1)\min_{z \in (\beta_1, \beta_2)}\left(-\frac{\partial }{\partial
z}h_{\bar{\alpha},d}(z)\right)
\min\left\{e^{-h_{\bar{\alpha},d}^2(\beta_1)},e^{-h_{\bar{\alpha},d}^2(\beta_2)}\right\}.\end{split}\eea
By definition (\ref{defchH}), for any $x,y$, $\frac{\partial }{\partial z} h_{x,y}(z)=\frac{\sigma}{z(z-1)\sqrt{2}}\leq-\sigma
2\sqrt{2}$; moreover,
\bea\min\left\{e^{-h_{\bar{\alpha},d}^2(\beta_1)},e^{-h_{\bar{\alpha},d}^2(\beta_2)}\right\}\geq\min\left\{e^{-h_{\bar{\alpha},d}
^2(\varepsilon)},e^{ -h_{\bar{\alpha},d}^2(1-\varepsilon)}\right\}=m_{\bar{\alpha},d}.\eea
Notice now that for any $d\in\Z$,  $m_{\bar{\alpha},d}\rightarrow 0$ if and only if $\alpha \to 1$; nevertheless, if $\alpha\to 1$, also
$(1+c_{\alpha})^{-1}\rightarrow 0$  and in particular there will be some $\alpha$ such that $(1+c_{\alpha})^{-1}<\ep$, which contradicts
the hypothesis $\beta_1\geq\ep$. Hence, can  we conclude that
\bea P\big((\alpha,d),(\beta_1,\beta_2)\times\{d\}\big)\geq\sigma\sqrt{2/\pi}~\min_{a}m_{\bar{\alpha},d}(\beta_2-\beta_1)>0\eea
where the minimum has to be computed for $\alpha$ satisfying the initial hypotheses.
\item If $\alpha=1$,$(\beta_1,\beta_2)\in[\ep,1-\ep]$ or if $\alpha \in(0,1)$, 
$(\beta_1,\beta_2)\subset[\frac{c_{\alpha}}{1+c_{\alpha}},1-\ep]\subseteq[\frac{2}{3},1-\ep]$:  by analogous procedure, we obtain \bea
P\big((\alpha,d),(\beta_1,\beta_2)\times\{d\}\big)\geq \sigma\sqrt{2/\pi}~\min_{\alpha}m_{\alpha,d}(\beta_2-\beta_1)>0\eea
where $m_{\alpha,d}=\min\left\{e^{-h_{\alpha,d}^2(\varepsilon)},e^{-h_{\alpha,d}^2(1-\varepsilon)}\right\}>0$ and its minimum is
computed for $\alpha$ satisfying the above hypotheses. The positiveness holds since for any $d\in\Z$,  $m_{\bar{\alpha},d}\rightarrow 0$
if and only if $\alpha \rightarrow 0$, which implies $\frac{c_{\alpha}}{1+c_{\alpha}}\rightarrow 1$ and contradicts $\beta_2\leq 1-\ep$.
\item Otherwise:
it is straightforward to verify that \bea P\big((\alpha,d),(\beta_1,\beta_2)\times\{d\}\big)\geq
\sigma\sqrt{2/\pi}~\left(m_{\alpha,d}+m_{\bar{\alpha},d}\right)(\beta_2-\beta_1).\eea
\end{enumerate}
Finally, if we consider
\begin{equation}
\bar{m}(\alpha,d,\beta_1,\beta_2)=\left\{\begin{array}{l}
m_{\bar{\alpha},d}\text{ if } \alpha=0 \text{ or if } \alpha \in (0,1) \text{ and }\varepsilon<\beta_1<\beta_2 \leq
\frac{1}{1+c_{\alpha}};\\
m_{\alpha,d}\text{ if } \alpha=1 \text{ or if } \alpha \in (0,1) \text{ and
}\frac{c_{\alpha}}{1+c_{\alpha}}<\beta_1<\beta_2\leq1-\varepsilon;\\
m_{\alpha,d}+m_{\bar{\alpha},d}\text{ otherwise. }\\
\end{array}\right.
\end{equation}
and
\begin{equation}
C^{(1)}_{\ep,d}=\sigma\sqrt{\frac{2}{\pi}}~\min_{{\alpha \in [0,1]}\atop{(\beta_1,\beta_2)\subset[\ep,1-\ep]}} \bar{m}
(\alpha,d,\beta_1,\beta_2)
\end{equation}
we conclude that for any $\ep>0$, $d\in\Z$,
\begin{equation}\label{concludethat}
 P\big((\alpha,d),(\beta_1,\beta_2)\times\{d\}\big)\geq C^{(1)}_{\varepsilon,d}(\beta_2-\beta_1)~~~~~~ C^{(1)}_{\varepsilon,d}>0.
\end{equation}

Let us prove the second inequality, on the basis of equations (\ref{sum_up2}). In this case, the component $d$ of the state moves to
$d+1$, which is not always possible in one step. In particular, there are two situations in which the transition probability is null:
$\alpha=1$ and when $\beta_1=\frac{c_{\alpha}}{1+c_{\alpha}}$ (and given the continuity of (\ref{sum_up2}, problems occur whenever
$\alpha\rightarrow 1$ or $\beta_1\rightarrow \frac{c_{\alpha}}{1+c_{\alpha}}$).

Both issues can be solved considering two-step transition: roughly speaking, if $\alpha$ is close to 1,  a first step is used to move
$\alpha$ away from 1 (and $d$ remains constant); at this point, the probability to move $d$ to $d+1$ is positive. On the other hand,
when  $\beta_1$ is close to $\frac{c_{\alpha}}{1+c_{\alpha}}$ a first step is used to move $d$ to $d+1$ and a second one to move the
component $\alpha$ to the desired interval (and now this is possible since we recast in the case in which $d$ remains constant,
previously studied).

 Let us assess this qualitative argumentation.

\begin{enumerate}
 \item
 If $\alpha=0$, $(\beta_1,\beta_2)\in[\ep,1-\ep]$ or if $\alpha \in(0,1-\delta_1]$ for some small $\delta_1>0$, 
$(\beta_1,\beta_2)\subset[\ep,\frac{c_{\alpha}}{1+c_{\alpha}}]$:
\be\label{servedopo} P\big((\alpha,d),(\beta_1,\beta_2)\times\{d+1\}\big)\geq
\sigma\sqrt{2/\pi}~ \min_{\alpha \in [0,1-\delta_1]}m_{\bar{\alpha},d+1}(\beta_2-\beta_1)>0\ee
where the positiveness of $\min_{\alpha \in [0,1-\delta_1]}m_{\bar{\alpha},d+1}>0$ as been discussed above.
\item If $\alpha \in (0,1-\delta_1]$, $\beta_1\in [\ep,\frac{c_{\alpha}}{1+c_{\alpha}}-\delta_2]$ for some small $\delta_1, \delta_2>0$
and $\beta_2\in[\frac{c_{\alpha}}{1+c_{\alpha}},1-\ep]$: the transition probability depends on $\beta_1$, not on $\beta_2$, and
\bea\begin{split} P\big((\alpha,d),(\beta_1,\beta_2)\times\{d+1\}\big)&\geq
\sigma\sqrt{2/\pi}~ \min_{\alpha \in
(0,1-\delta_1]}m_{\bar{\alpha},d+1}\left(\frac{c_{\alpha}}{1+c_{\alpha}}-\beta_1\right)\end{split}\eea
where $\frac{c_{\alpha}}{1+c_{\alpha}}-\beta_1\geq \delta_2\geq \delta_2(\beta_2-\beta_1)$.\\

Let us now consider the cases that require two steps to move with non-null probability into the desired set. For this purpose, notice
that 
\begin{equation}
  \begin{split}
 P^2&\big((\alpha,d),(\beta_1,\beta_2)\times\{d+1\}\big)=\\
&=\int_0^1\sum_{d'=d,d+1} P\big((\alpha,d),(\mathrm d\alpha',d')\big)P\big((\alpha',d'),(\beta_1,\beta_2)\times\{d+1\}\big)\\   
  \end{split}
  \end{equation}

\item If $\alpha \in (0,1-\delta_1]$, $\beta_1\in(\frac{c_{\alpha}}{1+c_{\alpha}}-\delta_2,\beta_2)$ and
$\beta_2\in[\frac{c_{\alpha}}{1+c_{\alpha}}, 1-\ep]$, we exploit that
\be
\begin{split}
P^2&\big((\alpha,d),(\beta_1,\beta_2)\times\{d+1\}\big)\geq\\
&\geq\int_0^1P\big( (\alpha,d),(\mathrm d\alpha',d+1)\big)P\big((\alpha',d+1),(\beta_1,\beta_2)\times\{d+1\}\big)  
\end{split}\ee
As $ P\big((\alpha',d+1),(\beta_1,\beta_2)\times\{d+1\}\big)\geq C^{(1)}_{\ep,d+1}(\beta_2-\beta_1)$ by (\ref{concludethat}),
\be\begin{split} P^2 &\big((\alpha,d),(\beta_1,\beta_2)\times\{d+1\}\big)\geq C^{(1)}_{\ep,d+1}(\beta_2-\beta_1)
P\big((\alpha,d),([0,1],d+1)\big)\\
&\geq C^{(1)}_{\ep,d+1}(\beta_2-\beta_1) P\big((\alpha,d),([\ep,1-\ep],d+1)\big)\geq\\
&\geq C^{(1)}_{\ep,d+1}(\beta_2-\beta_1)\sigma\sqrt{2/\pi}(1-2\ep)\min_{\alpha\in(0,1-\delta_1]}m_{\bar{\alpha},d+1}.\\
\end{split}\ee
\item If $\alpha \in (1-\delta_1,1]$, we exploit that
\be
\begin{split}
P^2&\big((\alpha,d),(\beta_1,\beta_2)\times\{d+1\}\big)\geq\\
&\geq\int_0^1 P\big((\alpha,d),(\mathrm d\alpha',d)\big)P\big((\alpha',d),(\beta_1,\beta_2)\times\{d+1\}\big).  
\end{split}\ee 
A sufficient condition to have $P\big((\alpha',d),(\beta_1,\beta_2)\times\{d+1\}\big)>0$ is $\beta_2\leq
\frac{c_{\alpha'}}{1+c_{\alpha'}}$ (see \ref{servedopo}) which corresponds to
$\alpha'^2-\alpha'+\exp\left(\frac{1}{\sigma^2}\right)\left(\frac{1-\beta_2}{\beta_2}\right)^2\geq 0$. This holds for any $\alpha'$ if
$4\exp\left(\frac{1}{\sigma^2}\right)\left(\frac{1-\beta_2}{\beta_2}\right)^2>1$, otherwise for
$\alpha'\in[0,\widetilde{\alpha}]\cup[1-\widetilde{\alpha},1]$ where
$\widetilde{\alpha}=\frac{1-\sqrt{1-4\exp\left(\frac{1}{\sigma^2}\right)\left(\frac{1-\beta_2}{\beta_2}\right)^2}}{2}$

Reducing the domain of integration to $[0,\widetilde{\alpha}]$, we obtain
\begin{equation}\begin{split}
P^2&\big((\alpha,d),(\beta_1,\beta_2)\times\{d+1\}\big)\geq\\
&\geq\int_0^{\widetilde{\alpha}} P\big((\alpha,d),(\mathrm d\alpha',d)\big)P\big((\alpha',d),(\beta_1,\beta_2)\times\{d+1\}\big)\\
&\geq\int_0^{ \widetilde{\alpha} } P\big( (\alpha, d),(\mathrm d\alpha',d)\big)\sigma\sqrt{2/\pi}~m_{\bar{\alpha'},d+1}
(\beta_2-\beta_1)\\
&\geq\sigma\sqrt{2/\pi}~\min_{\alpha'\in[0,\widetilde{\alpha}]}m_{\bar{\alpha'},d+1}(\beta_2-\beta_1)
P\big((\alpha,d),([0,\widetilde{\alpha}],d)\big)\\
&\geq\sigma\sqrt{2/\pi}~\min_{\alpha'\in[0,\widetilde{\alpha}]}m_{\bar{\alpha'},d+1}(\beta_2-\beta_1)C^{(1)}_{\ep,d}\widetilde{\alpha}.
\end{split}\end{equation}
\end{enumerate}
Finally, gathering the bounds obtained in the previous four cases, we obtain 
\begin{equation}
 P^2\big((\alpha,d),(\beta_1,\beta_2)\times\{d+1\}\big)\geq C^{(2)}_{\varepsilon,d}(\beta_2-\beta_1).
\end{equation}
where $C^{(2)}_{\ep,d}=\delta_2(1-2\ep)\widetilde{\alpha}\sigma\sqrt{2/\pi}\min_{\alpha\in[0,1-\delta_1]}m_{\bar{\alpha},d+1}\min
\{C^{(1)}_{\ep,d},C^{(1)}_{\ep,d+1}\}>0$.

We omit the proof of the third inequality as it is analogous to the second one: by the same argumentation, we obtain a suitable constant
$C_{\ep,d}^{(3)}$. Finally, for any small $\ep>0$ and $d\in\Z$, $C_{\ep,d}=\min\{C_{\ep,d}^{(1)}, C_{\ep,d}^{(2)}, C_{\ep,d}^{(3)}\}$.

The thesis is now proved for any open interval in $[\ep,1-\ep]$. The generalization to all the open sets in $[\ep,1-\ep]$ is
straightforward since any open set on the real line is countable union of disjoint open intervals. Finally, we can extend the result to
all the Borelians in $[\ep,1-\ep]$. Remind that for any Lebesgue measurable set $M$ (in particular, for any Borelian) in $\R$ there
exists a sequence of open sets $O_n$ such that $M\subset \cap_{n=1}^{\infty}O_n$ and
$\mathcal{L}(M)=\mathcal{L}(\cap_{n=1}^{\infty}O_n)$, see \cite{rud:66}. As any finite intersection of open sets is open, we have
\bea P^{r}\big((\alpha,d),\cap_{n=1}^{N}O_n\times\{d'\}\big)\geq C_{\ep}\mathcal{L}(\cap_{n=1}^{N}O_n)\geq
C_{\ep}\mathcal{L}(\cap_{n=1}^{\infty}O_n)=C_{\ep}\mathcal{L}(M)\eea for any $d'\in \{d-1,d,d+1\}$ and $r=1,2$ according to the value of
$d'$. This inequality holds for any $N \in \N$, hence \bea
\lim_{N\to\infty}P^{r}\big((\alpha,d),\cap_{n=1}^{N}O_n\times\{d'\}\big)=P^{r}((\alpha,d),\cap_{n=1}^{\infty}O_n\times\{d'\})\geq
C_{\ep}\mathcal{L}(M).\eea
\qed\end{proof}
By this lemma, it follows in particular that for any $M \in \FB([\ep,1-\ep])$,
\bea
\left\{\begin{array}{ll}
P^{2|d-d'|}\big((\alpha,d),M\times\{d'\}\big)\geq C_{\ep,d}^{|d-d'|}\mathcal{L}(M)&~\text{ if } d\neq d';\\ 
P\big((\alpha,d),M\times\{d\}\big)\geq C_{\ep,d}\mathcal{L}(M).\\  
\end{array}\right.
\eea 
Moreover,
\begin{proposition}\label{lower_bound} For any $M \in \FB([0,1])$ with $\mathcal{L}(M)>0$,
\bea
\left\{\begin{array}{ll}
P^{2|d-d'|}\big((\alpha,d),M\times\{d'\}\big)>\frac{1}{2} C_{\ep,d}^{|d-d'|}\mathcal{L}(M)&~\text{ if } d\neq d';\\ 
P\big((\alpha,d),M\times\{d\}\big)>\frac{1}{2} C_{\ep,d}\mathcal{L}(M).\\  
\end{array}\right.
\eea 
In particular, $(A_k,D_k)_{k \in \N}$ is $(\mathcal{L}\times \kappa)$-irreducible, $\kappa$ being the counting measure.
\end{proposition}
\begin{proof}
By the previous lemma, this result holds when $M \in \FB([\ep,1-\ep])$ given any $\ep>0$. Now, if we consider any $M\in\FB([0,1])$ with
$\mathcal{L}(M)=\lambda>0$, we have $\mathcal{L}(M\cap[\ep,1-\ep])=\mathcal{L}(M)-\mathcal{L}(M\cap[\ep,1-\ep]^\mathrm c)\geq
\lambda-2\ep$ and we can always choose $\ep=\ep(\lambda)$ such that $\lambda>2\ep$. For instance, let us choose $\ep=\frac{\lambda}{4}$,
so that $\lambda-2\ep=\frac{\lambda}{2}$. Therefore,  \bea \begin{split}P^{2|d-d'|}\big((\alpha,d),M\times\{d'\}\big)&\geq
P^{2|d-d'|}\big((\alpha,d),(M\cap[\ep,1-\ep])\times\{d'\}\big)\\
&\geq C_{\ep,d}\mathcal{L}(M\cap[\ep,1-\ep])>\frac{\lambda}{2}C_{\ep,d}^{|d-d'|}\end{split}\eea when $d\neq d'$, and similarly when
$d=d'$.
\qed\end{proof}

\subsection{Two States Algorithm: an upper bound for the transition probability kernel}\label{sec_ub}
\begin{lemma}\label{upper_bound}
There exists a real positive constant $G$ such that
\bea P\big((\alpha,d),M\times\Z\big)\leq G \mathcal{L}(M)\eea
for any $(\alpha,d) \in [0,1]\times \Z$ and $M \in \FB([0,1])$.
\end{lemma}
\begin{proof}
First, we prove the lemma when $M$ is an open interval. Consider the equations (\ref{sum_up1}) - (\ref{sum_up3}): given $(\alpha,d)$,
$P\big((\alpha,d),(\beta_1,\beta_2)\times\Z\big)$ is equal to a sum of
integrals of type $\int_{\beta_1}^{\beta_2}e^{-h_{x,y}^2(z)}(-h_{x,y}'(z))\mathrm d z$ with $x=\alpha,1/(1-\alpha)$ and $y=d-1,d,d+1$
according to the instance. As we have shown in the Proof of Lemma 2, $h_{x,y}'(z)=\frac{\sigma}{z(z-1)\sqrt{2}}$, hence
$g(z)=-e^{-h_{x,y}^2(z)}h_{x,y}'(z)>0$ for every $z \in (0,1)$. Furthermore, $g'(z)$ is monotone decreasing over $(0,1)$ and null in one
point $z_0\in(0,1)$ corresponding to the unique solution of the equation $h_{x,y}(z)=\frac{\sqrt{2}}{\sigma}(\frac{1}{2}-z)$; hence
$g(z)$ is increasing in $(0,z_0)$, decreasing in $(z_0,1)$ and admits a maximum in $z_0\in(0,1)$. In conclusion,
$\int_{\beta_1}^{\beta_2}g(z)\mathrm d z\leq G(\beta_2-\beta_1)$, $G=g(z_0)$.\\
The extension to all the open sets is trivial as any open set is countable union of disjoint intervals. Finally, as for any
$M\in\FB([0,1])$ there exists a sequence of open sets $O_n$ such that $M\subset \cap_{n=1}^{\infty}O_n$ and
$\mathcal{L}(M)=\mathcal{L}(\cap_{n=1}^{\infty}O_n)$ (see \cite{rud:66}), for any $n \in \N$ we can write
\bea P\big((\alpha,d),\cap_{n=1}^{\infty}O_n\times\Z\big)\leq P\big((\alpha,d),\cap_{n=1}^{N}O_n\times\Z\big)\leq
G\mathcal{L}(\cap_{n=1}^{N}O_n)\eea
as any finite intersection of open sets is open. The result follows from the arbitrariness of $N$. \qed\end{proof}
\subsection{Proof of Theorem \ref{strong2}}\label{par_proof2}
The process $(A_k,D_k)_{k \in \N}$ with  $ (U_k)_{k \in \N}$ is an instance of MCRE.
 The corresponding EMP in $\Omega=[0,1]\times \Z\times \{0,1\}^{\N}$ is defined by the following transition probability kernel:
\be\label{def_tpk_emp}
P\big((\alpha,d,\su),A\times\{d'\}\times
B\big)=P_{u_0}\big((\alpha,d), A\times\{d'\}\big)\mathds{1}_B
(T\su)\ee
where $\su=(u_0,u_1,\dots) \in \{0,1\}^{\N}$,  $A\in\FB([0,1])$, $d' \in \Z$, $B \in \mathcal{P}(\{0,1\}^{\N})$.

 $P_{u_0}\big((\alpha,d), A\times\{d'\}\big)$ can be assessed by equations (\ref{d0})-(\ref{dm1}).
 Moreover, we denote by $P_{u_0,\dots u_{k-1}}\big((\alpha,d), A\times \{d'\}\big)$ the probability of moving from $(\alpha,d)\in
[0,1]\times\Z$ to the set $A\times \{d'\}$, $A\in\FB([0,1])$, in $k$-steps, given the input sequence $(u_0,\dots,u_{k-1})\in \{0,1\}^k$.
By Proposition \ref{invariant_MCRE},
$\widetilde{\psi}=\widetilde{\phi}\times\pi$ ($\widetilde{\phi}$ being defined in Proposition \ref{admits_invariant}), is an invariant
p.m. for the EMP. Moreover,
\begin {lemma}
$\widetilde{\psi}$ is ergodic.
\end {lemma}
\begin{proof}
Let $F\subset\Omega$ be an invariant set: by Definition \ref{def_erg}, to prove the ergodicity of
$\widetilde{\psi}$ is sufficient to show that  $\widetilde{\psi}(F)>0$ implies  $\widetilde{\psi}(F)=1$. 

Then, let us suppose  $\widetilde{\psi}(F)>0$. We name
\bea
\begin{split}
&\mathcal{U}_F=\big\{\su\in\{0,1\}^{\N} : (\alpha,d,\su)\in F \text{ for some }(\alpha,d)\in [0,1]\times \Z\big\};\\
&\mathcal{U}_0=\big\{\su\in\{0,1\}^{\N} : \su \text{ contains infinitely many 0's and 1's}\big\};\\
&\mathcal{U}_0^{n}=\big\{\su\in\mathcal{U}_0 : \su \text{ contains at least a 0 and a 1 in its first $n$ bits }\big\},~n\geq 2.\\
\end{split}
\eea
Given the transition probability kernel (\ref{def_tpk_emp}), if $\su \in \mathcal{U}_F$ then also $T\su\in\mathcal{U}_F$
and since $\pi$ is an ergodic measure with respect to the shift operator $T$ (see \cite[Section 1.5]{wal:2000}) and
$\pi(\mathcal{U}_F)>0$ (otherwise  $\widetilde{\psi}(F)=0$), we have that 
$\pi(\mathcal{U}_F)=1$ by the Birkhoff's Individual Ergodic Theorem (\cite[Theorem 1.14]{wal:2000}).

By analogous reasoning, $\pi(\mathcal{U}_0)=1$. Furthermore,  $\mathcal{U}_0^n\subset \mathcal{U}_0^{n+1}$, then
$\mathcal{U}_0^n\uparrow \mathcal{U}_0$. This implies the existence of an $n_0\geq 2$ such that $\pi(\mathcal{U}_0^{n_0})>0$.

At this point, let us consider the equations (\ref{d0})-(\ref{dm1}): by applying the procedure used to prove Lemma
\ref{lontano_dai_bordi} and Proposition \ref{lower_bound}, it is easy to verify that for any $(\alpha,d)\in (0,1)\times \Z$,
\begin{equation}\label{cond_prob_lb}
  \begin{split}
    &P_{0}\big((\alpha,d),M\times\{d\}\big)>0~\text{ for any } M\in\FB\left(~\left(1/3,1\right]~\right),~\mathcal{L}(M)>0;\\
    &P_{1}\big((\alpha,d),M\times\{d\}\big)>0~\text{ for any } M\in\FB\left(~\left[0,2/3\right)~\right),~\mathcal{L}(M)>0;\\
    &P_{0}\big((\alpha,d),M\times\{d+1\}\big)>0~\text{ for any } M\in\FB\left(~ \left[0,2/3\right)~ \right),~\mathcal{L}(M)>0;\\
    &P_{1}\big((\alpha,d),M\times\{d-1\}\big)>0~\text{ for any } M\in\FB\left(~\left(1/3,1\right]~\right),~\mathcal{L}(M)>0.\end{split}
\end{equation}
where $\frac{1}{3}$ and $\frac{2}{3}$ are sufficient, not necessary bounds derived from Remark (\ref{calphabounds}). These inequalities
yield to
\be
\label{cond_prob_lb2}
  \begin{split}
& P_{01}\big((\alpha,d),M\times\{d\}\big)>0~\text{ for any } M\in\FB\left(~\left[0,1\right]~\right),~\mathcal{L}(M)>0;\\
& P_{10}\big((\alpha,d),M\times\{d\}\big)>0~\text{ for any } M\in\FB\left(~\left[0,1\right]~\right),~\mathcal{L}(M)>0.\\
\end{split}
\ee
Notice also that we are not considering the negligible cases $\alpha=0$ and $\alpha=1$, which may prevent the one-step transition (see
(\ref{d0})-(\ref{dm1})). Maintaining this hypothesis, consider  $(\alpha,d,\su)\in F$ such that  $\su\in \mathcal{U}_0^{n_0}$ (this is
always possible since $\mathcal{U}_0^{n_0}\subset \mathcal{U}_F$ $\widetilde{\psi}$-a.e.). By the invariance of $F$ and
(\ref{cond_prob_lb2}), we obtain that
\be\label{fibre} [0,1]\times \{d\}\times \{T^{n_0}\su\}\subset F \ee 
since $\su$ contains at least a 0 and a 1 in its first $n_0$ bits. Moreover, the fact that $\mathcal{U}_0^{n_0}$ is not negligible
implies that we can always choose $\su\in\mathcal{U}_0^{n_0}$ such that $\mathcal{V}_{\su}=\{T^n\su,~n\in\N\}$ has measure
$\pi(\mathcal{V}_{\su})=1$, as a consequence of \cite[Theorem 1.14]{wal:2000}. Hence,
\be\label{fibre2} [0,1]\times \{d\}\times \mathcal{V}_{\su}\subset F \ee 

Birkhoff.

Furthermore, consider the evolution of the component $d \in\Z$: from equations (\ref{cond_prob_lb}) we deduce that any $d$ has non-null
probability to achieve, in $n$ steps, any integer belonging to
\bea
D_n=\{d-m_1,d-m_1+2,\dots,d+n-m_1\} 
\eea
$m_1$ being the number of 1's in the corresponding $n$-bit input sequence. Hence,
\be\label{fibre3} [0,1]\times D_n \times T^n\mathcal{V}_{\su}\subset F \ee 
where $ T^n\mathcal{V}_{\su}= \mathcal{V}_{\su}$ $\pi$-a.e..
Given that for any $n$, $D_n\subset D_{n+1}$, in particular, $D_{n+1}$ has one more element than $D_n$, then $D_n\uparrow \Z$. This
finally proves that
\be\label{fibre4} [0,1]\times \Z \times \mathcal{V}_{\su}\subset F~~~\pi\text{-a.e.} \ee 

$\widetilde{\phi }(F_{\mathbf w}=1$. But now also $ [0,1]\times \Z \times \{\mathbf{T w}\}\subseteq F$.
which implies\be \widetilde{\psi}(F)=\widetilde{\phi }([0,1]\times \Z)\pi(\mathcal{V}_{\su})=1.\ee
\qed\end{proof}

Given $q(\alpha,d,U_k)=P(\UE_k \neq
U_k|U_k,A_k=\alpha,D_k=d)$,
\be \begin{split}&P_b(e|\RU)=\frac{1}{K}\sum_{k=0}^{K-1}P(\UE_k \neq
U_k|\RU)=\\ &=\int_0^1\sum_{d\in
\Z}\frac{1}{K}\sum_{k=0}^{K-1}q(\alpha,d,U_k)P_{(U_0,\dots U_{k-1})}\big((1,0), (\mathrm d\alpha,d)\big).\end{split}\ee

Now, let $g(\alpha,d,\RU)=q(\alpha,d,U_0)$: it is easy to verify that \bea P^k g(\alpha,d,\RU)=\int_0^1\sum_{d'\in
\Z}q(\alpha',d',U_k)P_{(U_0,\dots U_{k-1})}\big((\alpha,d), (\mathrm d\alpha',d')\big)\eea  then
\be P_b(e|\RU)=\frac{1}{K}\sum_{k=0}^{K-1} P^k g(1,0,\RU).\ee
 By the Ergodic Theorem \ref{Ergodic_T}, \bea \lim_{K \to \infty}\frac{1}{K}\sum_{k=0}^{K-1}(P^k g)(\omega)=\int_{\Omega}g~\mathrm
d\widetilde{\psi}\;\;\;\text{for }\; \widetilde{\psi}\text{-a.e. } \omega \in \Omega\eea
 Let $N\subset \Omega$ be the negligible set for which there is no convergence and let $N_{0,\RU}=\{\alpha \in [0,1]: (\alpha,0,\RU)\in
N\}$. By the same argumentation used in Corollary \ref{mean2}, $P_{U_0}((1,0),N_{u,\RU}\times\{0\})=0$ and 
\bea\begin{split}&P_b(e|\RU)
=\frac{1}{K}g(1,0,\RU)+\frac{1}{K}\sum_{k=1}^{K-1}\int_{\alpha_1\in[0,1]}P_{U_0}((1,0),(\mathrm d \alpha_1,0))
(P^{k-1}g)(\alpha_1,0,T\RU)\\
&\stackrel{K\to\infty}{\longrightarrow}\int_{\alpha_1\in[0,1]\setminus N_{0,\RU}}P_{U_0}((1,0),(\mathrm d \alpha_1,0))\int_{\Omega}
g~\mathrm d \widetilde\psi=\int_{\Omega} g~\mathrm d \widetilde\psi~~~\pi\text{-a.e.}\RU\in\{0,1\}^{\N}.
\end{split}\eea which proves the thesis, as \bea \int_{\Omega}g~\mathrm d \widetilde{\psi}=\int_{[0,1]}\sum_{d\in\Z}\sum_{u\in\{0,1\}}
q(\alpha,d,u)\widetilde{\phi}(\mathrm d\alpha,d)\pi_0(u)=\int_{[0,1]\times\Z} \qmad ~ \mathrm d \widetilde \phi.\eea

\end{document}